\documentclass{aa}
\usepackage{psfig}

\begin{document}

\title{Nonlinear stability of relativistic sheared planar jets}

\subtitle{}

\author{M. Perucho\inst{1}
          \and
        J. M. Mart\'{\i}\inst{1}
          \and
        M. Hanasz\inst{2}
}
 \offprints{M. Perucho}

   \institute{Departament d'Astronomia i Astrof\'{\i}sica,
              Universitat de Val\`encia, 46100 Burjassot (SPAIN)\\
   \email{manuel.perucho@uv.es,
              jose-maria.marti@uv.es} \and Toru\'n Centre for
              Astronomy, Nicholas Copernicus University, PL-97-148
              Piwnice k.Torunia (POLAND)\\
   \email{mhanasz@astri.uni.torun.pl}}

\date{Received .../ Accepted ...}

\abstract{The linear and non-linear stability of sheared, relativistic
planar jets is studied by means of linear stability analysis and
numerical hydrodynamical simulations. Our results extend the previous
Kelvin-Hemlholtz stability studies for relativistic, planar jets in
the vortex sheet approximation performed by Perucho et al. (2004a,b)
by including a shear layer between the jet and the external medium and
more general perturbations. The models considered span a wide range of
Lorentz factors ($2.5-20$) and internal energies
($0.08\,c^2-60\,c^2$) and are classified into three classes
according to the main characteristics of their long-term, non-linear
evolution. We observe a clear separation of these three groups in a
relativistic Mach-number Lorentz-factor plane. Jets with a low Lorentz
factor and small relativistic Mach number are disrupted after
saturation. Those with a large Lorentz factor and large relativistic
Mach number are the stablest, due to the appearance of short
wavelength resonant modes which generate local mixing and heating in
the shear layer around a fast, unmixed core, giving a plausible
solution for the problem of the long-term stability of relativistic
jets. A third group is present between them, including jets with
intermediate values of Lorentz factor and relativistic Mach number,
which are disrupted by a slow process of mixing favored by an
efficient and continuous conversion of kinetic into internal
energy. In the long term, all the models develop a distinct
transversal structure (shear/transition layers) as a consequence of KH
perturbation growth, depending on the class they belong to. The
properties of these shear layers are analyzed in connection with the
parameters of the original jet models.

 }

\titlerunning{}
\authorrunning{M. Perucho et al.}
\titlerunning{Nonlinear stability of relativistic sheared planar jets}

\maketitle

\section{Introduction}

  Transversal structure in extragalactic jets could be the natural
consequence of current formation mechanisms (see, e.g., Sol et
al. 1989; Celotti \& Blandford 2000), in which an ultrarelativistic
presumably electron/positron outflow from the high latitude region
close to the spinning black hole (and powered by, e.g., the extraction
of energy from the hole) is surrounded by a mildly relativistic,
electron/proton, hydromagnetic outflow launched from the outer parts
of the accretion disk. Recent numerical simulations of jet formation
from black hole magnetospheres (Koide et al. 1997) also lead to
outflows with {\it two-layered shell structure} consisting of inner,
fast gas pressure driven jets surrounded by slower, magnetically
dominated winds. On larger scales, shear layers (with distinct
kinematical properties and magnetic field configurations) have been
invoked in the past by several authors (Komissarov 1990, Laing 1996,
Laing \& Bridle 2002a,b) in order to account for a number of the
observational characteristics of FR~I radio sources. The model of De
Young (1993) to explain the FRI/FRII morphological dichotomy is based
on deceleration of the jet flow at the inner galactic core and the
subsequent formation of turbulent shear layers in FRIs. Recently,
Swain et al. (1998) found evidence of shear layers in FR~II radio
galaxies (3C353), and Attridge et al. (1999) have inferred a
two-component structure in the parsec-scale jet of the source
1055+018. On the other hand, first simulations of radio emission from
three-dimensional relativistic jets with shear layers (Aloy et
al. 2000) allowed several observational trends in parsec and
kiloparsec jets to be interpreted: inhomogeneous distributions of
apparent speeds within radio knots (Biretta et al. 1995); rails of low
polarization along jets (as in 3C353; Swain et al. 1998); top/down jet
emission asymmetries in the blazar 1055+018 (Attridge et
al. 1999). Stawarz \& Ostrowski (2002) have studied the contribution
to the radiative jet output from turbulent shear layers in large-scale
jets.

  Given all the pieces of theoretical and observational evidence
concerning the ubiquity of shear layers in extragalactic jets, it
appears natural to analyze their influence on the
dynamics and stability of relativistic jets. An attempt to investigate
the growth of the Kelvin-Helmholtz (KH) instability in some particular
class of sheared, cylindrical relativistic jets was pursued by
Birkinshaw (1991). However, this study is limited to the ordinary and
low-order reflection modes, and the domain of jet parameters
considered involves only marginally relativistic flows (beam flow
velocities $\leq 0.1 c$; $c$ is the speed of light) and
non-relativistic (jet, ambient) sound speeds ($\leq 0.01c$). Other
approaches to linear analysis of the stability of relativistic
stratified jets (Hanasz \& Sol 1996, Hardee \& Hughes 2003) and
sheared, ultrarelativistic jets (Urpin 2002) have also been taken.
Several recent works combine linear analysis and hydrodynamical
simulations in the context of both relativistic jets (Rosen et
al. 1999, Hardee 2000, 2001) and GRBs (Aloy et al. 2002).

  In this paper, we focus on the study of the evolution of
relativistic (planar) flows with shear layers through the linear and
non-linear regimes relying on both linear, analytical stability
analysis and hydrodynamical numerical simulations. The present work
complements the one presented in Perucho et al. (2004a,b; hereafter,
Paper~I and II, respectively), in which we characterized the effects of
relativistic dynamics and thermodynamics in the development of KH
instabilities in planar, relativistic jets in the vortex sheet
approximation. We used a more general setup for simulations with the
inclusion of a set of symmetric (pinching) and antisymmetric (helical)
sinusoidal perturbations in two dimensional slab jets and a thicker
shear layer ($\simeq 0.2\,R_j$) than that used to mimic vortex sheet
evolution. The use of slab jets allows for inclusion of helical
perturbations, which are known to be present in extragalactic
jets. Moreover, two dimensional simulations provide the possibility of
a much larger resolution than three dimensional ones. The aim of this
work is to study the stability properties of jets depending on their
initial parameters and the effect of shear layers in those
properties. We used the temporal approach, which allows for larger
resolution, and fixed two different grid sizes, depending on the
thermodynamical properties of jets, which are neither directly related nor
coupled to the wavelength of a specific mode, as was the case in
Papers I and II. Jet parameters are based on those of previous papers
for direct comparison. We generalized our results with simulations
where only one antisymmetric mode is perturbed (similar to
simulations in Paper I), and simulations of cylindrical jets, where several
modes are perturbed (as in simulations presented in this paper).

  The plan of this paper is as follows. In Sect.~\ref{sect:numsim}
we describe the numerical simulations and present the parameters used
in this paper. In Sect.~\ref{sect:results} we describe our results
concerning linear and nonlinear regimes of new simulations; we
discuss them in Sect.~\ref{sect:disc} and present our conclusions in Sect.~\ref{sect:concl}.

\section{Setup for numerical simulations}
\label{sect:numsim}

%
\begin{table*}
\begin{center}
$
\begin{array}{|c|cc|ccccccc|cccc|}
\hline
 {\rm Model} &        \gamma &    \varepsilon_j &    \varepsilon_a &       c_{sj}  &        c_{sa} &
    p &           \nu &          \eta &           M_j  & k_0& k_1& k_2& k_3\\
\hline
         {\rm A2.5} &            2.5 &          0.08 &         0.008 &          0.18 &    0.059 &     0.0027 &      0.11 &            0.11 &     12.5   & 0.39 & 0.78 & 1.18 & 1.57 \\
         {\rm B2.5} &            2.5 &          0.42 &         0.042 &          0.35 &    0.133 &     0.014 &       0.14 &            0.15 &     6.12   & 0.39 & 0.78 & 1.18 & 1.57  \\
         {\rm D2.5} &           2.5 &          60.0 &          6.000 &          0.57 &    0.544 &     2.000 &       0.87 &            0.90 &     3.29   & 0.78 & 1.57 & 2.36 & 3.14  \\
         {\rm B05} &            5 &            0.42 &          0.042 &          0.35 &    0.133 &     0.014 &       0.14 &            0.15 &     13.2   & 0.39 & 0.78 & 1.18 & 1.57  \\
         {\rm D05} &            5 &            60.0 &          6.000 &          0.57 &    0.544 &     2.000 &       0.87 &            0.90 &     7.01   & 0.78 & 1.57 & 2.36 & 3.14 \\
         {\rm A10} &           10 &            0.08 &          0.008 &          0.18 &    0.059 &     0.0027&       0.11 &            0.11 &     54.2    & 0.39 & 0.78 & 1.18 & 1.57 \\
         {\rm B10} &           10 &          0.42 &          0.042 &            0.35 &    0.133 &     0.014 &       0.14 &            0.15 &     26.9   & 0.39 & 0.78 & 1.18 & 1.57  \\
         {\rm D10} &           10 &         60.0 &          6.000 &            0.57 &     0.544 &     2.000 &       0.87 &            0.90 &     14.2   & 0.78 & 1.57 & 2.36 & 3.14 \\
         {\rm B20} &           20 &          0.42 &          0.042 &           0.35 &     0.133 &     0.014 &       0.14 &            0.15 &     54.0  & 0.39 & 0.78 & 1.18 & 1.57  \\
         {\rm D20} &           20 &         60.0 &          6.000 &            0.57 &     0.544 &     2.000 &       0.87 &            0.90 &     28.5  & 0.78 & 1.57 & 2.36 & 3.14 \\

\hline
\end{array}
$
\end{center}

\caption{Equilibrium parameters of different simulated jet models.
The meaning of the symbols is as follows. $\gamma$: jet flow Lorentz
factor; $\varepsilon$: specific internal energy; $c_s$: sound speed;
$p$: pressure; $\nu$: jet-to-ambient relativistic rest mass density
contrast; $\eta$: jet-to-ambient enthalpy contrast; $M_j$: jet
relativistic Mach number; $k_{0,1,2,3}$: excited longitudinal
wavenumbers. Labels $a$ and $j$ refer to ambient medium and jet,
respectively. All the quantities in the table are expressed in units
of the ambient density $\rho_{0a}$, the speed of light $c$, and the
jet radius $R_j$.}

\label{tab:param}
\end{table*}

%

  The equations governing the evolution of a
relativistic perfect-fluid jet (see Paper I) are
\begin{equation}
\gamma^2 \!\!\left( \rho \!+\! {p\over c^2} \right)\!\! \left[\!
{\partial \vec v \over \partial t}\!+\! (\vec v  \cdotp \! \vec\nabla)
\vec v \! \right]
\! + \vec\nabla p + {\vec v \over c^2} {\partial p \over \partial t}
\! = \! 0, \label{eulereq}
\end{equation}
\begin{equation}
\gamma \!\!\left( {\partial \rho\over \partial t} \!+\! \vec v \cdotp
\vec \nabla\rho \right) \!+\! \left( \rho + {p\over c^2} \right)\!\!
\left[ {\partial \gamma \over \partial t} + \vec\nabla \cdotp (\gamma \vec v)
\right] \!\!= \! 0, \label{conteq}
\end{equation}
\begin{equation}
\frac{d(p \rho_0^{-\Gamma})}{dt} = 0.
\label{eq:adiabat}
\end{equation}
In the preceding equations, $c$ is the speed of light, $\rho_0$ the
particle rest mass density. $\rho$ stands for the relativistic density
which is related to the particle rest mass density and the specific
internal energy $\varepsilon$, by $\rho =
\rho_0(1+\varepsilon/c^2)$. The enthalpy is defined as $w = \rho +
p/c^2$, and the sound speed is given by $c_s^2 = \Gamma p/w$, where
$\Gamma$ is the adiabatic index. The relation between pressure and the
specific internal energy is $p =(\Gamma - 1) \varepsilon \rho_0$. The
velocity of the fluid is represented by $\vec v$, and $\gamma$ is the
corresponding Lorentz factor. The operator $d\,\,/dt$ appearing in
Eq.~(\ref{eq:adiabat}) is the standard Lagrangian time derivative.

  The steady initial flow is 2D-planar and symmetric with respect to
the $x=0$ plane. The flow moves in the positive $z$ direction.  Our
simulations were performed in the so-called temporal approach, as in
Papers I and II. In this approach, the evolution of perturbations in a
peridiocal slice of an infinite jet is followed along the time. In
order to study the effects of shearing in the development of the
instability, we assumed a continuous transition between the jet and the
ambient (the same as the one considered by Ferrari et al. 1982). The
profiles of the axial velocity and proper rest-mass density across
this transition layer, $v_z(x)$ and $\rho_0(x)$, respectively, are given by
\begin{equation}\label{sv}
    v_z(x)=\frac{v_{z,j}}{\cosh (x/R_j)^m},
\end{equation}
\begin{equation}\label{sr}
    \rho_0(x)=\rho_{0,a}-\frac{\rho_{0,a}-\rho_{0,j}}{\cosh (x/R_j)^m}.
\end{equation}
In the previous expressions, $v_{z,j}$ represents the fluid flow
velocity in the jet axis, whereas $\rho_{0,j}$ and $\rho_{0,a}$
are the proper rest-mass density at the jet axis and in the
ambient, respectively. The exponent $m$ controls the shear layer
steepness; in the limit $m \rightarrow \infty$ the configuration
tends to the vortex-sheet case. In our present calculations we
have used $m = 25$, corresponding to a shear layer of thickness
$0.17\,R_j$, about twice that used in Papers I and II, in order
to mimic the vortex sheet limit. From now on all quantities
representing the jet will be assigned the '$j$' subscript and
the quantities representing the ambient medium will be assigned
the '$a$'.

  Following conclusions given in the Appendices of both Paper I
and Paper II, the numerical resolution used was
$256\,{\rm cells}/R_j$ in the transversal direction times $32 \,{\rm
cells}/R_j$ in the axial direction. Note that we reduced the
transversal resolution with respect to the simulations in Papers I and
II. One reason for that were computational time limitations, as now our
grids are twice as large in the transversal direction as those used in
Paper I, since we are now combining symmetric and antisymmetric
structures. However, in the present simulations, transversal
resolution is not as critical as in the previous works, since we are
not interested in mimicking the evolution of instability in the
vortex sheet limit and therefore do not have steep shear
layers. The saving in transversal resolution allowed us to double
axial resolution, which affects the non-linear results (see Appendix
in Paper II). The physical sizes of grids are $8\,R_j$ axially times
$6\,R_j$ transversally for hot jets (models D, see Table
\ref{tab:param}) and $16\,R_j$ axially times $6\,R_j$ transversally
for cold jets (models A and B in Table \ref{tab:param}). The different axial
size is due to hot models having shorter unstable modes; see Sect.
\ref{sect:lin}, where we show linear problem solutions for one cold
and one hot jet.

Previous to these simulations, we performed several
representative runs (B05, D05, B20, D20) with the aim of studying
the evolution of models under single eigenmode perturbations in
planar antisymmetric geometry, exciting the first
reflection antisymmetric mode at its peak growth rate. These simulations
were used to check the consistency of the numerical results in the
linear phase with the linear stability analysis for relativistic,
sheared flows. Discussion of the evolution of these models
through the non-linear regime can be found in Appendix A.
Tests were also performed in order to
assess the difference in the evolution of linear and non-linear
regimes using a general sinusoidal perturbation (to be used in this
paper) and a superposition of eigenmodes, as done with the first body
mode alone in Paper I. Results showed that structures and qualitative
properties of the resulting flow were basically the same. This fact
confirms that general perturbations excite eigenmodes of the system.

  The parameters used in the simulations are shown in Table
\ref{tab:param}. We swept a wide range in Lorentz factors (from 2.5 to
20) and in specific internal jet energies (from $0.08 c^2$ to $60
c^2$) in order to obtain a global view of the response of different
initial data sets to perturbations. All the models correspond to a
single-component ideal gas with adiabatic exponent $\Gamma=4/3$.
These parameters were chosen in order to study the stability regions
found in Paper II: Class I for cold and slow models, which were deeply
mixed and mass loaded; Class II for hot and fast jets, which were
slowly mixed in the nonlinear regime, progressively losing their
axial momentum; Class III for hot and slow jets, with properties
between Classes I and II, and Class IV for cold/warm and fast models,
which were the stablest in the nonlinear regime. We performed
simulations for models \textbf{B05}\footnote{Boldface will be used for
new simulations in order to differentiate them from those in Paper II
with the same name.}, \textbf{B10}, \textbf{B20}, \textbf{D05},
\textbf{D10}, and \textbf{D20} of Paper II, and added
\textbf{A2.5} (same thermodynamical properties as A05 in Paper II),
\textbf{A10}, \textbf{B2.5}, and \textbf{D2.5}. Models \textbf{A2.5},
\textbf{B2.5}, and \textbf{B05} correspond to regions of Class I
jets. Models \textbf{D10} and \textbf{D20} correspond to Class II,
\textbf{D2.5} and \textbf{D05} to Class III, and
\textbf{A10},\textbf{B10} and \textbf{B20} belong to Class
IV.

  Perturbations we applied adding the following sinusoidal form to
transversal velocity, $v_x(x,z)$:
\begin{eqnarray}\label{perts}
\small
    v_{x} =\hspace{8cm}\nonumber{} \\
    \frac{V_{x1}}{N}
    \left(\sum_{n=0}^{N-1}
    \sin((n+1)\,k_{n}\,z\,+\,\varphi_n)  
    \sin^2((n+1)\,\pi\,x) \frac{x}{|x|}\right)+\nonumber{} \\
    \frac{V_{x1}}{M}
    \left(\sum_{m=0}^{M-1}
    \sin((m+1)\,k_{m}\,z\,+\,\varphi_m)\sin^2((m+1)\,\pi\,x)\right),\,\,\,\,
\end{eqnarray}
where $V_{x1}(\sim 10^{-4})$ is the amplitude given to the
perturbation, $k_{m,n}$ are the wavenumbers of the grid box (so
that $(n+1)\,k_n$ and $(m+1)\,k_m$ stand for the harmonics of the
symmetric (pinching) and antisymmetric (helical) modes,
respectively), and $\varphi_n$ and $\varphi_m$ are random phases
given to each mode. In our simulations, four symmetric ($N = 4$)
plus four antisymmetric modes ($M = 4$) were excited, i.e., the
fundamental mode of the box and the first three harmonics.

  Numerical simulations were performed using a finite-difference code
based on a high-resolution shock-capturing scheme which solves the
equations of relativistic hydrodynamics written in conservation form
(Mart\'{\i} et al. 1997 and references therein). Before
performing the simulations, several improvements were made in the
numerical code. In particular, the relativistic PPM routines
(Mart\'{\i} \& M\"uller 1996) were properly symmetrized. The code was
recently parallelized using OMP directives. Simulations were performed
with 4 processors on SGI 2000 and SGI Altix 3000 machines.

\section{Results}
\label{sect:results}

\subsection{Linear regime}
\label{sect:lin}

\subsubsection{Perturbation theory}

  We introduced an adiabatic perturbation of the form $\propto g(x)
\exp (i(k_z z - \omega t))$ to the flow equations
(\ref{eulereq}-\ref{eq:adiabat}), $\omega$ and $k_z$ ($k_x$) being the
frequency and wavenumber of the perturbation along (across) the jet
flow. We followed the temporal approach, in which perturbations
grow in time, with real wavenumbers and complex frequencies, with the
imaginary part being the growth rate. By linearizing the equations
and eliminating the perturbations of rest mass density and flow
velocity, a second-order ordinary differential equation for the
pressure perturbation $P_1$, is obtained (Birkinshaw 1991, Aloy et
al. 2002)
\begin{eqnarray}
  P_1^{\prime\prime} + \left(\frac{2\gamma_0^2v_{0z}^\prime (k_z - \omega
  v_{0z})}{\omega -v_{0z}k_z} - \frac{\rho_{e,0}^\prime}{\rho_{e,0} +
  P_0}\right)P_1^\prime  + & & \label{radial-eq}\\
  \gamma_0^2\left(\frac{(\omega  -v_{0z}k_z)^2}{c_{s,0}^2} - (k_z - \omega
  v_{0z})^2\right)P_1 & = & 0 \nonumber
\end{eqnarray}
where $\rho_{e,0}$ is the energy-density of the unperturbed model,
$P_0$ the pressure, $v_{0z}$ the three-velocity component,
$\gamma_0 = 1/\sqrt{1-v_{0z}^2}$ the Lorentz factor, and
$c_{s,0}$ the relativistic sound speed. The prime denotes the
$x$-derivative. Unlike the vortex sheet case, in the case of a
continuous velocity profile, a dispersion relation cannot be
written explicitly. The equation (\ref{radial-eq}) is integrated
from the jet axis, where boundary conditions on the amplitude of
pressure perturbation and its first derivative are imposed
\begin{eqnarray}
\label{eq:bcs1}
P_1(x = 0) = 1, & P_1^{\prime}(x = 0) = 0 & \mbox{(sym. modes)},
\\ \nonumber
P_1(x = 0) = 0, & P_1^{\prime}(x = 0) = 1  & \mbox{(antisym.
modes)}.
\end{eqnarray}
Solutions satisfying the Sommerfeld radiation conditions (no
incoming waves from infinity and wave amplitudes decaying towards
infinity) are found with the aid of the numerical method, based on the
shooting method (Press et al. 1997) proposed in Roy Choudhury \&
Lovelace (1984).

\begin{figure}[!h]
\centerline{\psfig{file=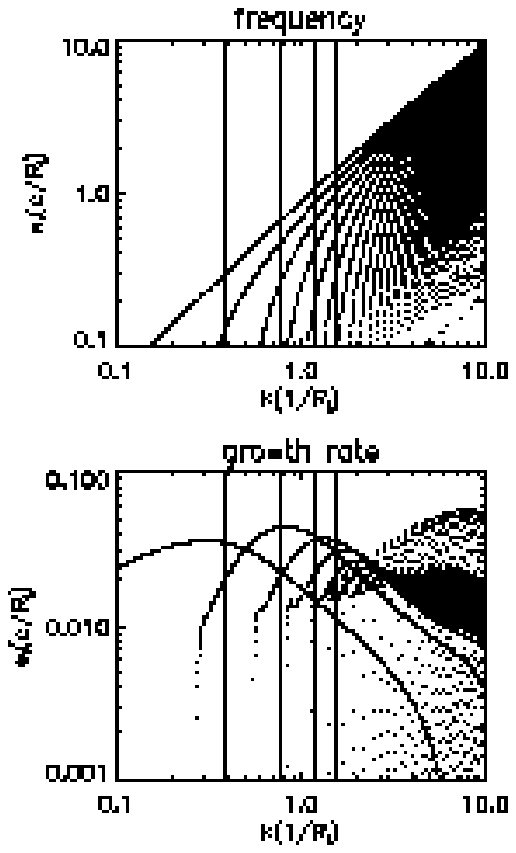,width=0.22\textwidth,angle=0,clip=}
\qquad \psfig{file=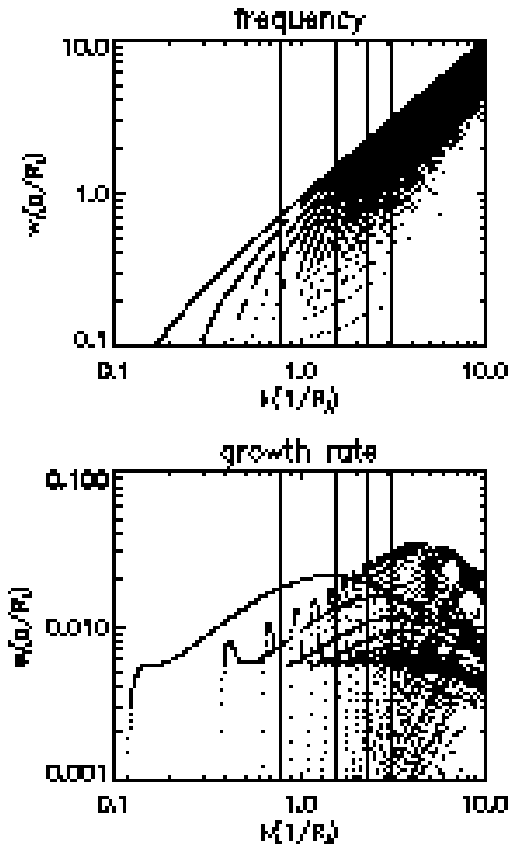,width=0.22\textwidth,angle=0,clip=}}
\caption{Solutions for the linear perturbation differential
equation. Left panel: Antisymmetric solution for model
\textbf{B05}. Right panel: Symmetric solution for model
\textbf{D20}. Vertical lines stand for the perturbed wavenumbers
in the numerical simulations (from left to right: $k_0$, $k_1$,
$k_2$ and $k_3$). Let us note that the fundamental mode does 
not appear in the right panel, as its growth rates are lower than the 
scale of the plot.} \label{fig:linsol}
\end{figure}

  Linear stability analysis was performed for all models presented in
Sect.~\ref{sect:numsim}, in the symmetric and antisymmetric
cases. Figure~\ref{fig:linsol} shows examples of solutions for the
linear problem for sheared jets in models \textbf{B05} and
\textbf{D20}. Top panels in Fig. \ref{fig:linsol} show the (real part
of) frequency as a function of wavenumber, and bottom panels show the
imaginary part of frequency or growth rate, defined as the inverse of
the time needed by a given mode to $e$-fold its amplitude, also as a
function of wavenumber. Each mode is defined by its wavelength,
frequency and growth rate.
%
%

\begin{figure}[!h]
\centerline{
\psfig{file=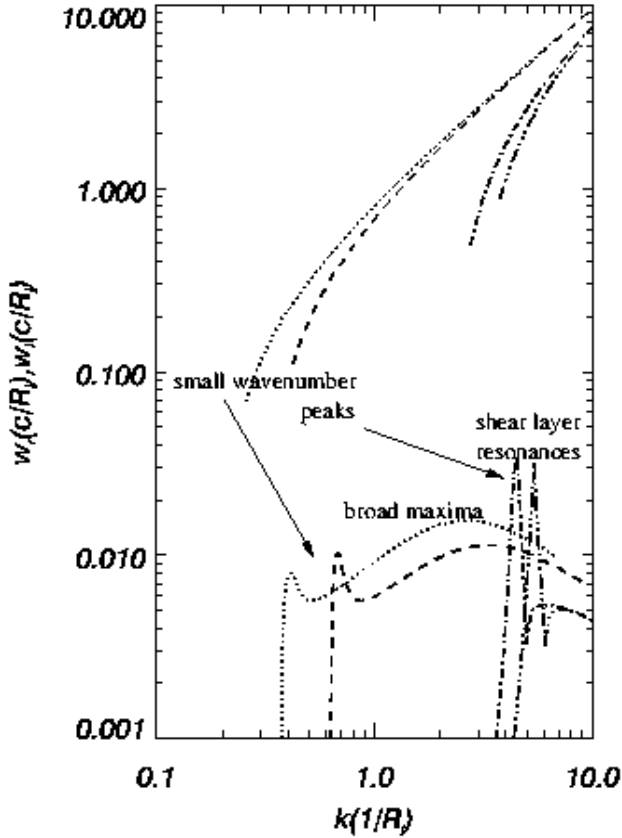,width=0.5\textwidth,angle=0,clip=}}
\caption{Specific modes of solutions shown in Fig.~\ref{fig:linsol},
symmetric solution for model \textbf{D20}. Dotted line: first body
mode; dashed: second body mode; dash-dot: twentieth body mode;
dash-triple dot: twenty-fifth body mode. Arows point to both the broad
maxima and the small wavenumber peaks present in every single
mode. Low wavenumber peaks of high order body modes show higher
growth rates and are thus defined as (shear layer) resonances.}
\label{fig:linsol2}
\end{figure}

  From these results, we note that the individual reflection mode
solutions of the shear problem present lower growth rates for most
wavenumbers, especially in the large wavenumber limit, than do the
corresponding solutions in the vortex sheet case. This behaviour was
noticed for the first time by Ferrari et al. (1982) for the
first and second reflection modes in the non-relativistic limit. The
growth rate curves corresponding to a single $n_x$-th reflection mode
consists of a broad maximum at higher wavenumbers and a local peak
which is placed in the low wavenumber limit, near the marginal
stability point of a chosen reflection mode. Regarding the
relativistic case, while in the vortex-sheet limit the small
wavenumber peaks for individual modes are relatively unimportant
(since the maximum growth rates at these peaks are lower than those of
other unstable modes), while in the presence of the shear-layer they display
high growth rates for high order body modes. Therefore we shall call
these peaks {\em the shear layer resonances}. In
Fig.~\ref{fig:linsol2} we show the solution for four specific modes of
model 
\textbf{D20}, from
Fig.~\ref{fig:linsol}. 
Low order body modes do not show high peaks
at maximum unstable wavelengths, whereas high order body modes
show peaks (shear layer resonances) at this maximum wavelength 
and do not present broad maxima.
%
%

  From Eq.~(\ref{radial-eq}) we see that radial structure of
perturbations depends on physical parameters of the flow, as well
as on the given frequency and axial wavenumber of a given
perturbation. Resonances are determined by this transversal
structure, and therefore we should expect changes in their
characteristics depending on the properties of the shear layer and
physical parameters: i) a decrease of the jet Lorentz factor
reduces the dominance of the growth rates of resonant modes
with respect to ordinary and low order reflection modes; ii) a
decrease in the specific internal energy of the jet causes
resonances to appear at longer wavelengths; iii) further
widening of the shear layer reduces the growth rates and the
dominance of the shear-layer resonances, suggesting that there is
an optimal width of the shear layer that maximizes the effect for
a given set of jet parameters; the largest growth rate of resonant
modes moves to smaller wavenumbers and lower order reflection
modes; iv) perturbations with wavenumber higher than some limiting
value (that decreases with the shear layer width) are
significantly diminished (short-wavelength cut-off), consistent with
previous non-relativistic results (Ferrari et al. 1982). The
discovery of the shear layer resonances and their potential role
in the long-term stability of relativistic jets is reported
in Perucho et al. (2005).

\subsubsection{Simulations}
 
\begin{figure*}[!t]
\centerline{
\psfig{file=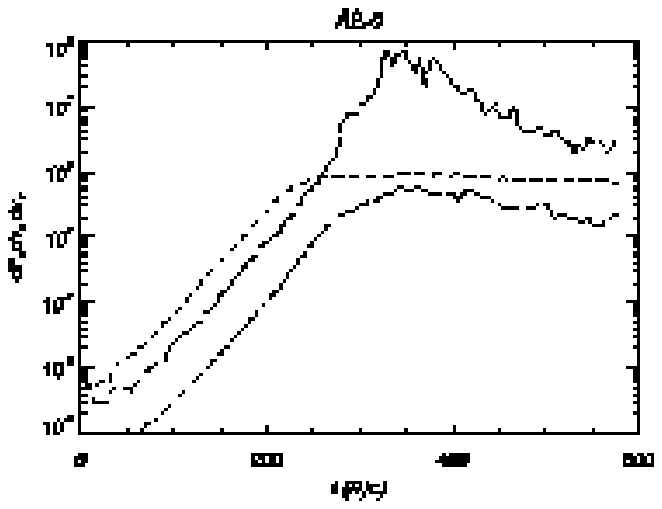,width=0.3\textwidth,angle=0,clip=}\quad
\psfig{file=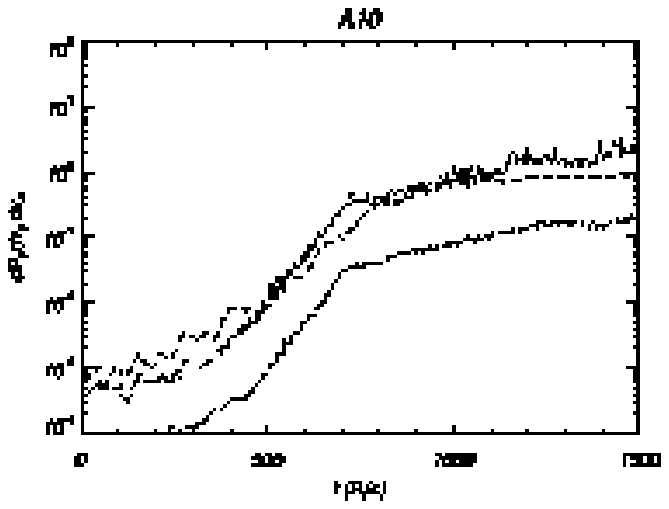,width=0.3\textwidth,angle=0,clip=}}
\centerline{
\psfig{file=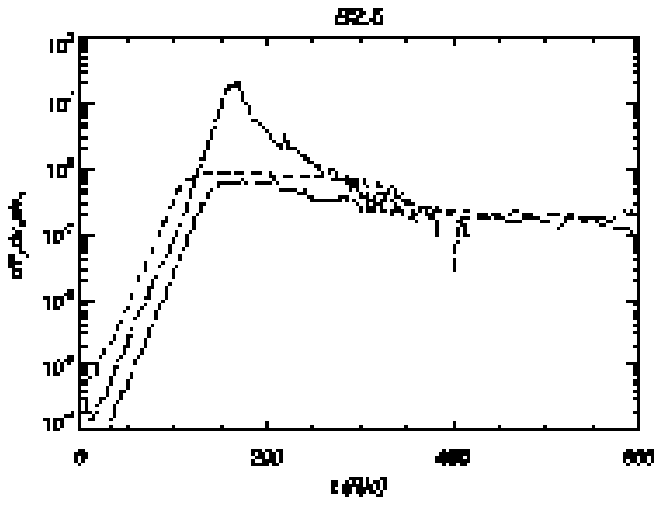,width=0.3\textwidth,angle=0,clip=}\quad
\psfig{file=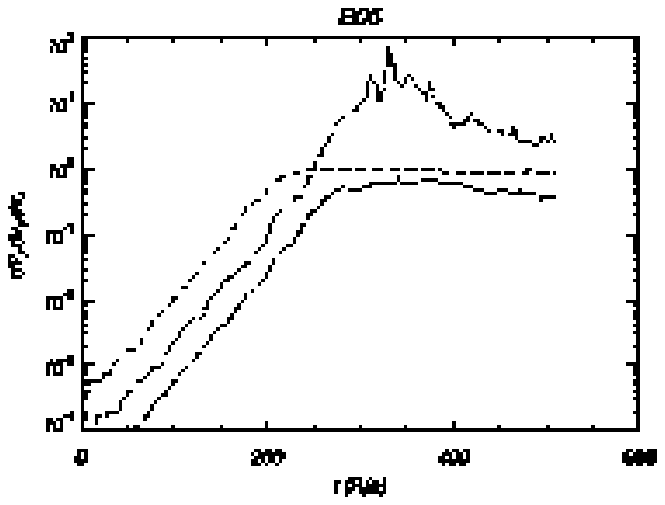,width=0.3\textwidth,angle=0,clip=}}

\centerline{
\psfig{file=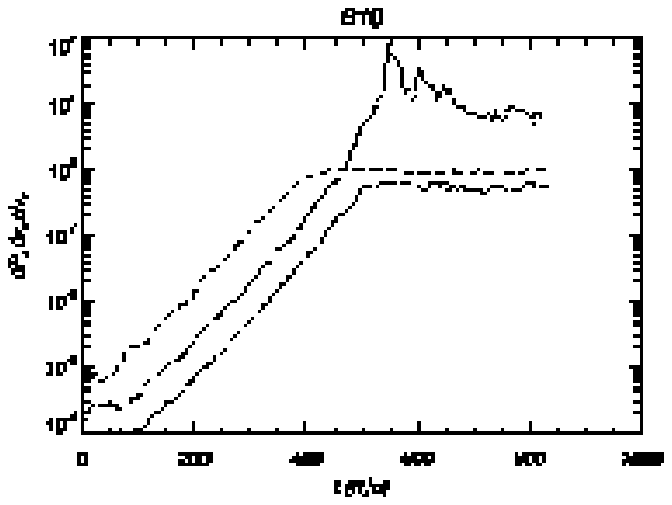,width=0.3\textwidth,angle=0,clip=} \quad
\psfig{file=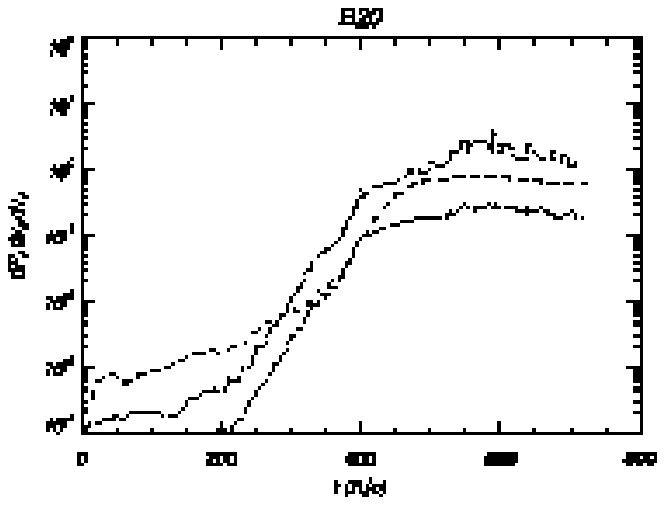,width=0.3\textwidth,angle=0,clip=}}
\centerline{
\psfig{file=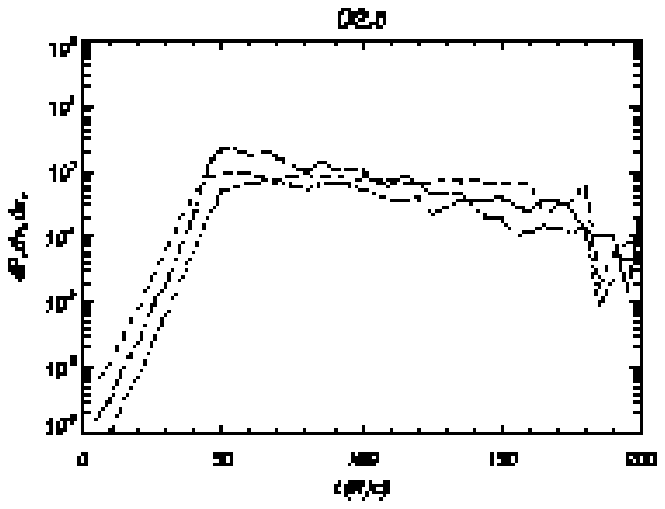,width=0.3\textwidth,angle=0,clip=} \quad
\psfig{file=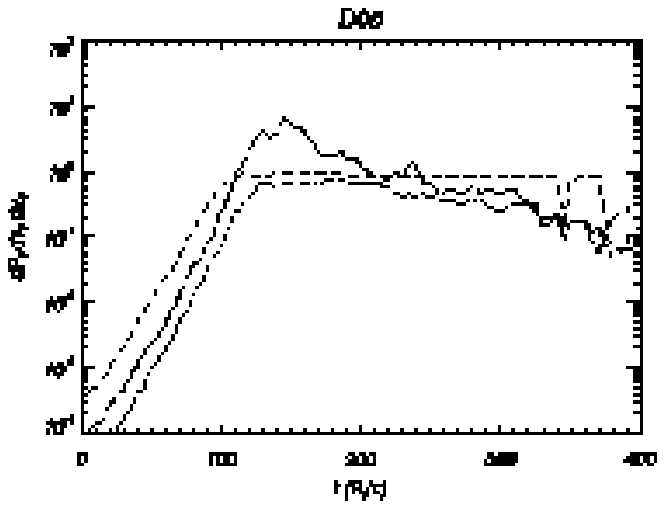,width=0.3\textwidth,angle=0,clip=}}
\centerline{
\psfig{file=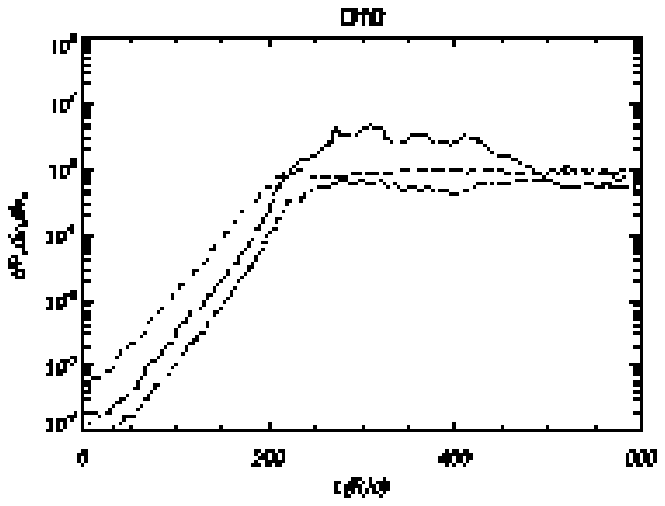,width=0.3\textwidth,angle=0,clip=} \quad
\psfig{file=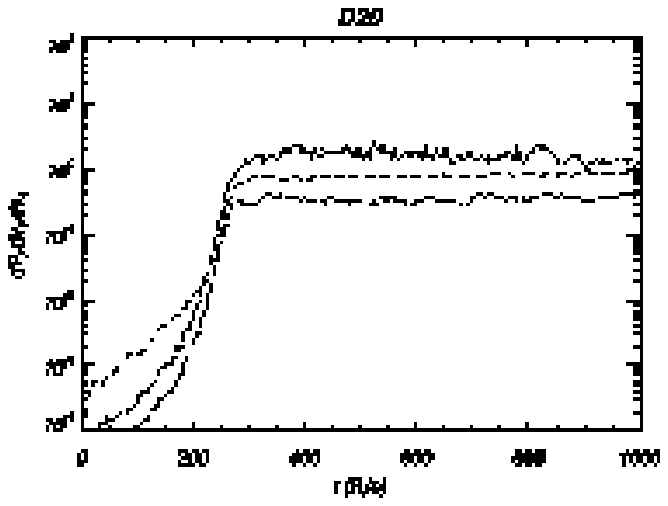,width=0.3\textwidth,angle=0,clip=}}

\caption{Evolution of the relative amplitudes of perturbations.
Dotted line: pressure perturbation ($(p_{max}-p_0)/p_0$). Dashed line:
longitudinal velocity perturbation in the jet reference frame
($0.5\,(v'_{z,max}-v'_{z,min})$). Dash-dotted line: perpendicular
velocity perturbation in the jet reference frame
($0.5\,(v'_{x,max}-v'_{x,min})$). Note the different scales in the
horizontal axis. The search for maximum ($p_{max}$, $v'_{x,max}$,
$v'_{z,max}$) and minimum ($v'_{x,min}$, $v'_{z,min}$) values were
restricted to those numerical zones with jet mass fraction larger
than 0.5.}
\label{fig:amplitudes}
\end{figure*}

Table~\ref{tab:t1} summarizes the properties of the linear phase in
our simulations. The left part of the Table (colums 2-9) gives the
theoretical growth rates of the perturbed wavelengths, taken at the
vertical lines in Fig.~\ref{fig:linsol}. The last column
gives the values of the growth rate corresponding to the dominant
wavelength as deduced from Fourier analysis of the transversal
profiles of the rest mass density distribution in the jet. Note,
however, that Fourier analysis can only give us information about
wavelengths, but cannot distinguish between symmetric and
antisymmetric modes. The growths of pressure, axial, and transversal
velocity perturbations along the simulations are shown in
Fig.~\ref{fig:amplitudes}.

%
\begin{table*}
\begin{center}
\begin{tabular}{|c|cccc|l|c|}
\hline
Model&$w_{i,k_0}$& $w_{i,k_1}$& $w_{i,k_2}$& $w_{i,k_3}$&Dominant& $w_{i}$\\
&Symm.$\quad $Antis.&Symm. $\quad $Antis.&Symm.$\quad $  Antis.&Symm. $\quad $ Antis.&&\\
 \hline
         {\rm A2.5}&0.036 $\quad $  0.032&0.038$\quad $ 0.037&0.034 $\quad $ 0.036&0.031$\quad $ 0.034& $k_0$&0.030 \\
         {\rm B2.5}&0.042 $\quad $  0.056&0.070$\quad $ 0.052&0.066 $\quad $ 0.084&0.073$\quad $ 0.080&$k_1$,$k_2$ &0.070 \\
         {\rm D2.5}&0.046 $\quad $  0.160&0.131$\quad $ 0.182&0.210 $\quad $ 0.194&0.142$\quad $ 0.256&$k_2$,$k_1$&0.200\\
         {\rm B05} &0.037 $\quad $  0.035&0.037$\quad $ 0.044&0.036 $\quad $ 0.038&0.034$\quad $ 0.035&$k_0$,$k_1$&0.035\\
         {\rm D05} &0.068 $\quad $  0.063&0.085$\quad $ 0.063&0.100 $\quad $ 0.068&0.068$\quad $ 0.110&$k_1$,$k_2$,$k_0$&0.080\\
         {\rm A10} &0.009 $\quad $  0.009&0.006$\quad $ 0.006&0.005 $\quad $ 0.006&0.006$\quad $ 0.006&$k_0$*&0.004 (0.005)\\
         {\rm B10} &0.022 $\quad $  0.018&0.019$\quad $ 0.021&0.018 $\quad $ 0.017&0.013$\quad $ 0.013&$k_0$&0.020\\
         {\rm D10} &0.034 $\quad $  0.038&0.041$\quad $ 0.037&0.044 $\quad $ 0.034&0.051$\quad $ 0.035& $k_1$,$k_2$&0.040\\
         {\rm B20} &0.011 $\quad $  0.010&0.009$\quad $ 0.010&0.007 $\quad $ 0.007&0.009$\quad $ 0.010& $k_0$*&0.006 (0.008)\\
         {\rm D20} &0.018 $\quad $  0.018&0.020$\quad $ 0.017&0.022 $\quad $ 0.017&0.027$\quad $ 0.028& $k_1$,$k_0$&0.016\\

\hline
\end{tabular}

\end{center}

\caption{Dominant modes in the linear phase of the numerical
  simulations.  $w_{i,k_j}$: maximum growth rate for the $j$-th
  overtone wavenumber excited in the simulation (see
  Table~\ref{tab:param}) derived from linear stability analysis.  Left
  columns: symmetric mode; right columns: antisymmetric one. The
  dominant mode refers to the mode with the largest amplitude in rest
  mass density perturbation as derived from Fourier analysis of the
  box, written from larger to smaller amplitude when more
  than one is present. $w_{i}$: fitted pressure perturbation growth
  rate for the linear regime in the simulation. Growth rate values are
  in $c/R_j$ units.  *: Models where irregular growth affects the
  evolution (see text).}

\label{tab:t1}
\end{table*}

%

%
%

  Comparison of the evolution through the linear phase of the
different models in the numerical simulations and from the linear
stability analysis is summarized as follows:

\begin{itemize}

  \item \textbf{A2.5}: modes with longer wavelengths are faster
growing, and their Fourier amplitudes are consistently larger than
those for shorter modes in the simulation. Growth rate found in the
simulation is close to the one expected from linear stability
analysis.

  \item \textbf{B2.5}: first ($k_1$) and second ($k_2$) harmonics of
the box have larger amplitudes in the Fourier analysis and, therefore,
dominate the linear regime. Linear stability analysis gives $k_1$,
$k_2$ and $k_3$ as the fastest growing modes, with a similar growth rate
to that found in the simulation. However, $k_3$ modes have smaller
amplitudes.

  \item \textbf{D2.5}: found growth rate for the simulation is close
to that of $k_1$ and $k_2$ modes, which is confirmed by Fourier
analysis.  Antisymmetric $k_3$ mode might grow with slower rates than
theory predicts due to numerical viscosity that affects shorter modes
more than longer ones.

  \item \textbf{B05}: Fourier analysis shows competition between
fundamental and first harmonics of the box ($k_0$ and $k_1$,
respectively). This, as well as the value of the mean growth rate, is
confirmed by the linear stability analysis. The second harmonic of the
box ($k_2$) is damped.

  \item \textbf{D05}: according to Fourier analysis, $k_1$ and $k_2$
modes dominate evolution in the linear regime. The growth rate is
close to that of the symmetric $k_1$ mode, despite the fact that
symmetric $k_2$ and antisymmetric $k_3$ present faster growth rates,
so they appear to be damped.

  \item \textbf{A10}: Fourier analysis shows that longer modes
dominate, in agreement with linear stability analysis. However, the
growth rate in the numerical simulation is two times smaller than
predicted. We also observe in Fourier analysis that very short modes,
excited as harmonics of perturbed wavelengths, become important by the
end of the linear regime.

  \item \textbf{B10}: $k_0$ modes dominate, as predicted by linear
stability analysis.

  \item \textbf{D10}: as in models \textbf{D2.5} and \textbf{D05},
$k_1$ and $k_2$ have larger amplitudes in Fourier analysis, but the
smaller wavelength modes ($k_3$) are damped with respect to the
predictions of linear stability analysis.

  \item \textbf{B20}: longer modes dominate the linear evolution, in
agreement with linear analysis, but the growth rate in the numerical
simulation is $1.5$ times smaller than predicted. After some time,
short, fast modes, like those appearing in model \textbf{A10}, become
dominant and lead to a smooth transition to the non-linear regime.

  \item \textbf{D20}: long modes present larger amplitudes with
predicted growth rates up to the moment when shorter modes reach
larger amplitudes, the same effect as found in models \textbf{A10} and
\textbf{B20}.

\end{itemize}

  It is observed in several simulations (e.g., \textbf{B2.5},
\textbf{B05}, \textbf{D2.5}, \textbf{D05}, \textbf{D10}) that modes with
similar or even slightly higher growth rates than those dominating in
simulations present smaller amplitudes in the linear regime. It happens
usually for shorter modes (typically $k_2$, $k_3$), so it may be
caused by numerical viscosity, for less cells are involved in one
wavelength. However, the way in which we perturb the jet may also
favor the dominating growth of certain modes starting with a larger
amplitude. We added a general sinusoidal perturbation, so the input
amplitude of the perturbation at a given wavelength is shared in a
random way among all the modes present at that wavelength. This makes
some modes start their growths with smaller amplitudes, as we could
see in the Fourier analysis of different models. Initial low
amplitudes are more probable for short wavelength modes, as more
eigenmodes are present at a given wavenumber in this range (see
Fig.~\ref{fig:linsol}). From an initial lower amplitude, and taking
into account that they have similar growth rates to other modes, they
grow with smaller amplitudes for the rest of the linear phase.

  Models \textbf{A10} and \textbf{B20}, marked with an asterisk in
Table~\ref{tab:t1}, have fitted growth rates in the first part of the
linear regime below the predicted values. Note that these models have
the lower growth rates. After this initial phase, short harmonics
start dominating the linear growth.

%
\begin{table} 
\begin{center}
\begin{tabular}{|c|ccc|}
\hline
Model&$w_{i,{\rm max}}$&$w_{i,p,v_\perp}$&$w_{i,v_\parallel}$\\
 \hline
 {\rm B05} & 0.052 & $-$   & $-$   \\
 {\rm D05} & 0.11  & $-$   & $-$   \\
 {\rm A10} & 0.013 & 0.017 & 0.009 \\
 {\rm B10} & 0.035 & $-$   & $-$   \\
 {\rm D10} & 0.057 & $-$   & $-$   \\
 {\rm B20} & 0.026 & 0.036 & 0.036 \\
 {\rm D20} & 0.035 & 0.070 & 0.047 \\

\hline
\end{tabular}

\end{center}

\caption{Growth of resonant modes. Models which present a global
maximum growth rate (according to the linear analysis) for all
resonant modes (i.e., at any wavelength) above the growth rates of
the perturbed modes are listed. $w_{i,max}$: maximum growth rate
for all resonant modes from linear analysis; $w_{i,p,v_{\perp}}$:
fitted growth rates of pressure and perpendicular velocity
perturbations for the fast growth linear regime in the simulation,
only for those simulations where it occurs; $w_{i,\parallel}$:
same as $w_{i,p,v_{\perp}}$ for axial velocity. All values are in
$c/R_j$ units.}

\label{tab:t2}
\end{table}
%
%
\begin{figure*}
\centerline{\psfig{file=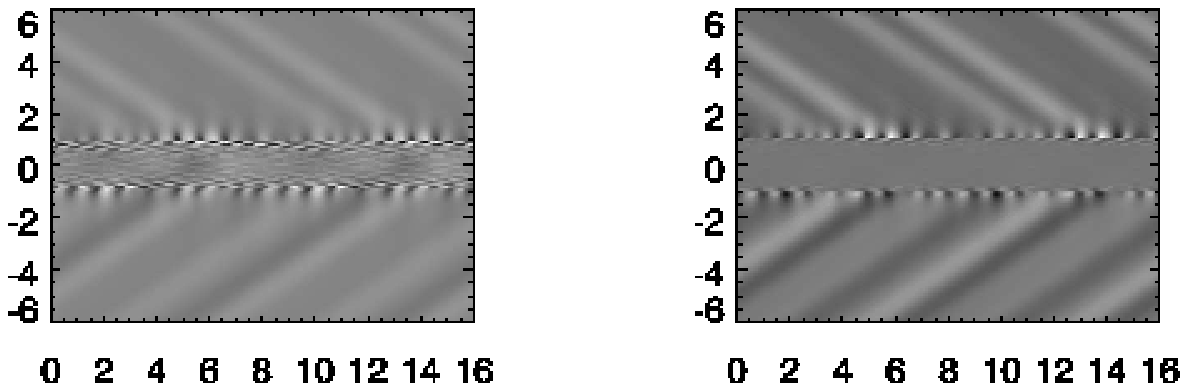,width=0.8\textwidth,angle=0,clip=}}
\centerline{\psfig{file=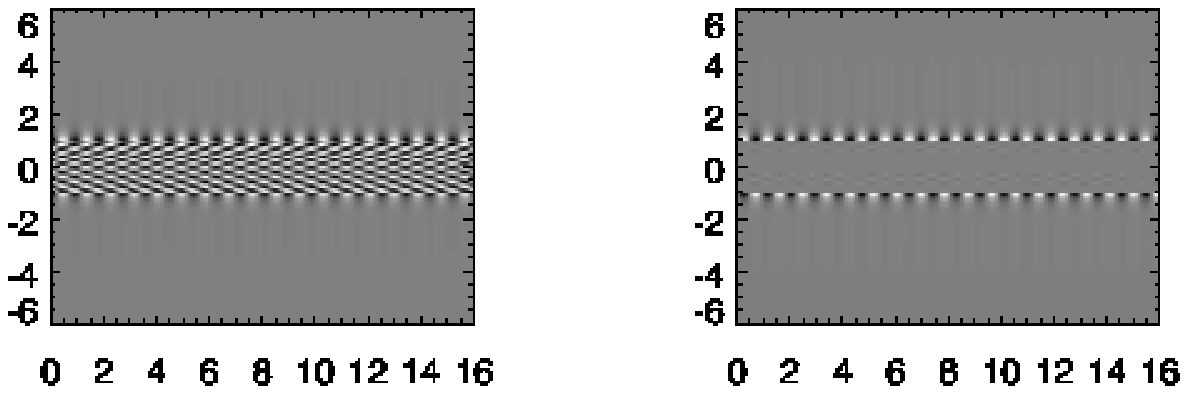,width=0.8\textwidth,angle=0,clip=}}
\caption{Upper panels: pressure (left) and perpendicular velocity
perturbation (right) at late stages of linear phase (model D20).
Lower panels: pressure (left) and perpendicular velocity
perturbation (right) corresponding to one resonant mode from
linear analysis. The linear, grey scale is arbitrary. Amplitudes
are maxima at the shear layer, hence the name of {\it shear layers
resonances} given to these modes. Oblique waves in upper panels
are the result of longer wavelength perturbations, not present in
the bottom panels.} \label{fig:restheo}
\end{figure*}
%

  We have observed the appearance of fast growing, very short modes in
models \textbf{A10}, \textbf{B20}, and \textbf{D20}, which are clearly
associated to the resonant modes presented above in the previous
section and which could have been excited as harmonics of the
initially perturbed wavelengths. The same kind of resonant mode might
have developed in model C20 of Paper II and caused the irregular
linear growth found with respect to the rest of models.  These modes
generate a rich internal structure in the jet due to their large
perpendicular wavenumber or, equivalently, short perpendicular
wavelengths (characteristic of high order body modes). A direct
comparison between the structure generated by these resonant modes in
the numerical simulations and that coming from linear stability
analysis can be seen in Fig.~\ref{fig:restheo}. In this figure, we
display one snapshot from model \textbf{D20} and the theoretical
counterpart using one of those resonant modes. The upper plots
correspond to the numerical simulation in the linear regime, where the
signature of the initial perturbations ($k_0$,$k_1$,$k_2$ and $k_3$)
are the oblique waves seen outside the jet. As seen in these plots,
resonant modes grow to amplitudes larger than those of the long waves,
as indicated by the black/white scale saturating precisely on the
interface. The lower plots represent the theoretical structure that we
would find if we had only excited a resonant mode and that is fairly
similar to the one appearing in the shear layer of the simulated
jet. Let us point out, however, that it is difficult to identify the
exact mode in the simulation, as the resonant modes overlap so much
(see Fig. 1) and it may happen that what we see is the structure
resulting from a combination of competing resonant modes.

According to the linear stability analysis, resonant modes have the
highest growth rates in high Lorentz factor jets and, among them, in
colder jets. This could be the reason why they only appear in
simulations of models \textbf{A10}, \textbf{B20}, and
\textbf{D20}. Table~\ref{tab:t2} collects the models which present
maximum growth rate for all resonant modes (i.e., at any wavelength),
found in the solutions to the linear problem, above the growth rates
of the perturbed modes. Maximum growth rates for resonant modes in
those models where they have been found, along with the fitted growth
rates in the simulation, are listed. Typically, the growth rates from
the numerical simulations are about $1.4-2.0$ times higher. This
difference remains unexplained, but it could be caused by second-order
effects, like interaction between modes.

  Summarizing, two kinds of linear growth are found in these
simulations, one dominated by longer modes typical of slower jets
and another one where short, fast modes appear. This difference is
important, for the transition to the non-linear evolution depends
critically on the dominant modes at the end of the linear regime.

   Table~\ref{tab:phases} shows the times at which linear phase
ends. As the end of the linear regime we selected the moment
when one of the variables (usually axial velocity) changes its slope
(departs from the linear growth, see definitions in Paper I). On the
other hand, we noticed that the longitudinal velocity perturbation
grows linearly up to values close to the speed of light and then beyond
the sound speed. This means that shocks should form at the end of
linear phase, as it is the case; see Fig. 3 in Paper I, where we
observe weak shocks starting to appear as conical structures. We
could have selected the end of the linear regime as the moment 
when these shocks start to appear. This would relate the end of the
linear regime directly to the internal sound
speed. We see that colder jets have longer linear phases than hot
ones, due to smaller typical growth rates in the former. $t_{lin}$
times are larger than those in Paper I, as growth rates are reduced by
the presence of the shear layer. Model \textbf{A10} presents the
longest linear phase.

%
\begin{table}
\begin{center}
$
\begin{array}{ccccccc}
\hline {\rm Model} & t_{\rm lin} & t_{\rm mex} & t_{\rm mix} &
t_{\rm sat} & t_{\rm peak} & \Delta_{\rm peak}
\\
\hline
{\rm A2.5} & 225 & 250 & 300 & 340 & 340 & 100  \\
{\rm B2.5} & 110 & 125 & 140 & 150 & 165 & 20 \\
{\rm D2.5} & 40 & 45 & 50 & 50 & 50 & 2 \\
\hline
{\rm B05} & 220 & 275 & 300 & 280 & 330 & 70  \\
{\rm D05} & 105 & 125 & 110 & 130 & 140 & 2 \\
\hline
{\rm A10} & 725 & - & - & - & - & 3 \\
{\rm B10} & 400 & 520 & 500 & 500 & 540 & 100  \\
{\rm D10} & 205 & 260 & 220 & 290 & 300 & 4  \\
\hline
{\rm B20} & 475 & - & 520 & 550 & 590 & 4  \\
{\rm D20} & 275 & 510 & 300 & 275 & 320 & 2 \\
\hline
\end{array}
$

\caption{Times for the different phases in the evolution of the
perturbed jet models. $t_{\rm lin}$: end of linear phase (the
amplitudes of the different quantities are not constant any
longer). $t_{\rm sat}$: end of saturation phase (the amplitude of the
transverse speed perturbation reaches its maximum). $t_{\rm mix}$: the
tracer starts to spread.  $t_{\rm peak}$: the peak in the amplitude of
the pressure perturbation is reached. $t_{\rm mex}$: the jet has
transferred to the ambient 1\% of its initial momentum. $\Delta_{\rm
peak}$: relative value of pressure oscillation amplitude 
at the peak of pressure perturbation (see Fig.~\ref{fig:amplitudes}).}
\label{tab:phases}

\end{center}
\end{table}
%

\subsection{Saturation and transition to nonlinear regime}
%
\begin{figure*}
\centerline{\psfig{file=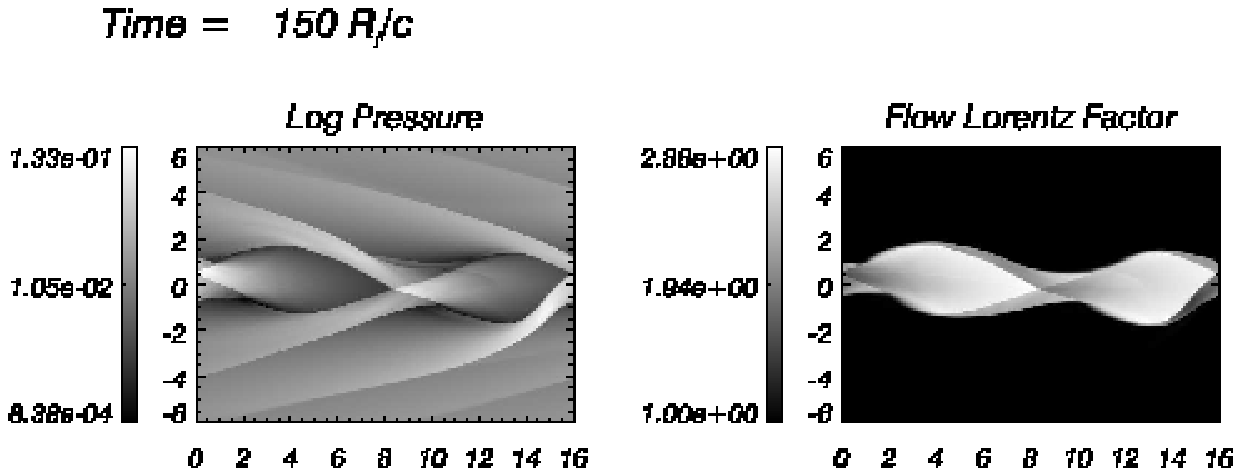,width=0.75\textwidth,angle=0,clip=}}
\centerline{\psfig{file=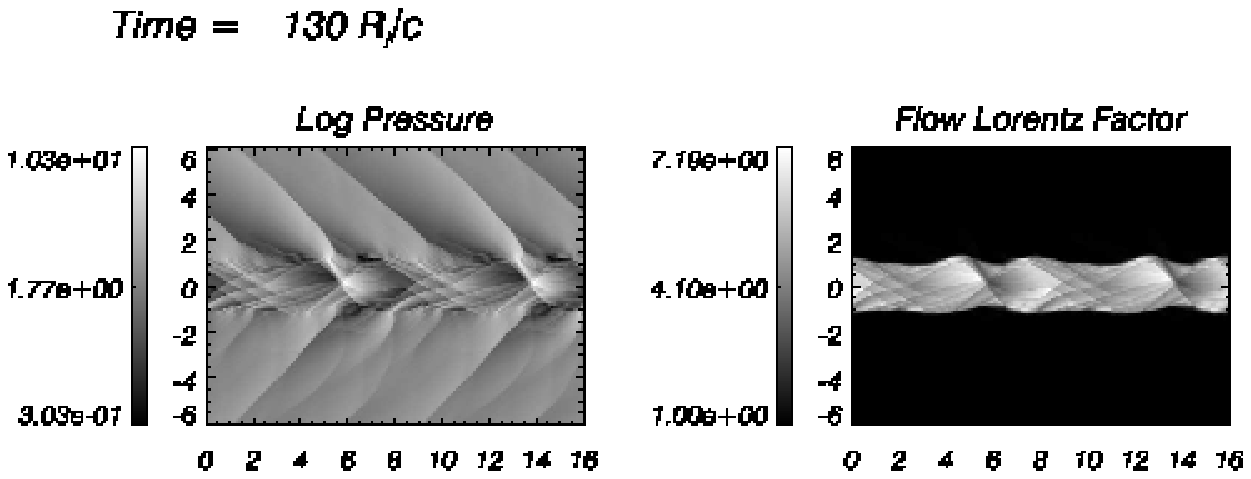,width=0.75\textwidth,angle=0,clip=}}
\centerline{\psfig{file=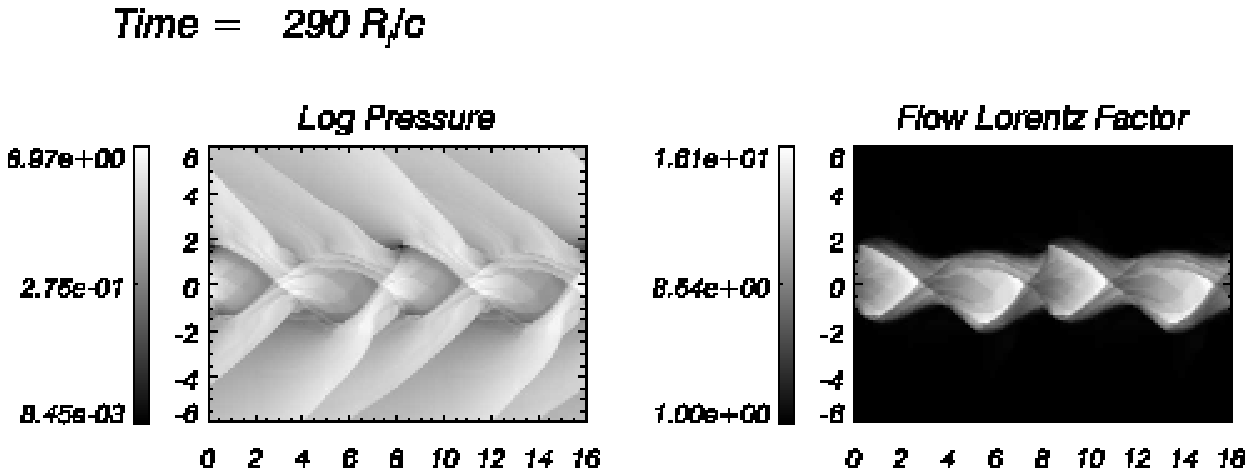,width=0.75\textwidth,angle=0,clip=}}

\caption{Snapshots of logarithm of pressure (left) and Lorentz factor
(right) for models \textbf{B2.5} (upper panels), \textbf{D05} (center
panels) and \textbf{D10} (bottom panels) at $t_{sat}$, where irregular
structures caused by mode competition are observed.}

\label{fig:fsat}
\end{figure*}
%

  Saturation of perturbations is reached (see Paper I) when
perpendicular velocity cannot grow further in the jet reference frame
due to the speed of light limit. Saturation times $t_{\rm sat}$ for
the different models are listed in Table~\ref{tab:phases}.  In this
phase, structures generated by dominating modes become visible in the
deformations of the jet. In Fig.~\ref{fig:fsat} we show snapshots of
three models (\textbf{B2.5}, \textbf{D05}, and \textbf{D10}) at
saturation time where mode competition derived from Fourier analysis
is clearly observed. Asymmetric structures appear as a result of
several symmetric and antisymmetric modes with large amplitudes.

%
\begin{figure*}[!t]

\centerline{
\psfig{file=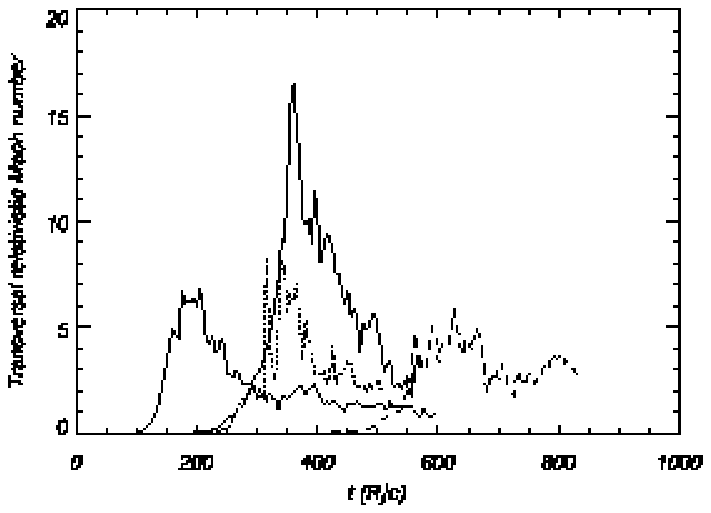,width=0.45\textwidth,angle=0,clip=} \quad
\psfig{file=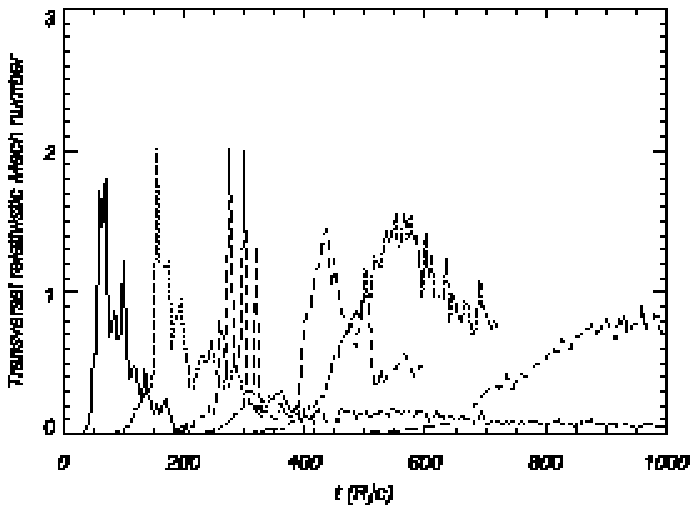,width=0.45\textwidth,angle=0,clip=}}

\caption{Transversal Mach number in simulations (see text for
definition). Solid line: $\gamma=2.5$; dotted line: $\gamma=5.0$;
dashed line: $\gamma=10.0$; dash-dot line: $\gamma=20.0$. The thinnest
line is for model A and thickest for model D, with B in
between, the left panel shows models \textbf{A2.5}, \textbf{B2.5},
\textbf{B05}, and \textbf{B10}, and the right panel shows models
\textbf{A10}, \textbf{B20}, \textbf{D2.5}, \textbf{D05}, \textbf{D10},
and \textbf{D20}.}

\label{fig:f2}
\end{figure*}
%
  In Paper II we also discussed how at the end of the saturation
phase nearly all the simulations lead to a sharp peak in the pressure
oscillation amplitude. These peaks are also seen in the present
simulations (see Fig.~\ref{fig:amplitudes}). The relative values of
pressure oscillation amplitude at the peak $\Delta_{\rm peak}$ and
the corresponding times $t_{\rm peak}$ are listed in
Table~\ref{tab:phases}. The values of $\Delta_{\rm peak}$ were
connected with the non-linear evolution of the flow. Those cases in
which $\Delta_{\rm peak} > 70$ developed a shock in the jet/ambient
interface followed by the sudden disruption of the jet. From
Table~\ref{tab:phases}, we see that peak values in the present
simulations are in general qualitatively the same as the corresponding
ones in Paper~I. Colder and slower jets have larger peaks and hence
suffer stronger shocks after saturation. The main difference between
the values in this paper and those presented in Paper~I appears for
models \textbf{B20} and \textbf{D20}, where shock strength is much
smaller due to the appearance of resonant, stabilizing modes, as we
discuss next.

  The parallel and perpendicular wavelengths of the shear-layer
resonant modes, $\lambda_z$ and $\lambda_x$, respectively, are both
small ($\lambda_z\leq R_j$) with $\lambda_x \ll \lambda_z$. Therefore
their wavevectors are almost perpendicular to the jet axis so the
waves propagate from the shear layer towards the jet axis. On the
other hand, the resonant modes have high growth rates, exceeding the
growth rate of other modes, so they start to dominate in the
evolution. Subsequently, the resonant modes saturate as soon as the
flow velocity oscillation amplitude approaches the speed of light. As
the maximum amplitude is reached, the sound waves steepen while
travelling towards the jet axis and form shock fronts on the leading
edges of wave profiles Dissipation of the oscillation energy of
resonant modes in shocks changes the background flow, so that the
amplification conditions of the longer wavelength modes change during
the course of time, reducing the value of $\Delta_{\rm peak}$ and
preventing the formation of a strong shock.

  Finally, as found in Paper II, the generation of the shock
wave at the jet/ambient interface is imprinted in the evolution of the
maxima of the transversal Mach number of the jet with respect to the
unperturbed ambient medium. This quantity is defined as
$M_{j,\perp}=\gamma_{j,\perp}v_{j,\perp}/(\gamma_{c_{sa}}c_{sa})$,
with $\gamma_{j,\perp}$ and $\gamma_{c_{sa}}$ the Lorentz factors
associated to $v_{j,\perp}$ and $c_{sa}$, respectively. A value
significantly larger than 1 around $t_{\rm peak}$ points towards a
supersonic expansion of the jet at the end of the saturation
phase. This magnitude is shown in Fig. \ref{fig:f2}. We observe a
clear inverse tendency of the peak value of this magnitude from colder
to hotter and from slower to faster jets, with the exception of
\textbf{A10} with respect to \textbf{B10} and \textbf{D10}, due to the
presence of the resonant stabilizing modes preventing the formation of
a strong shock. 

It is important to note that models with $\Delta_{\rm peak}>10$
(\textbf{A2.5}, \textbf{B2.5}, \textbf{B05}, and \textbf{B10}) coincide
with those developing larger transversal Mach numbers, see top panels
of Fig. \ref{fig:fsat} for model \textbf{B2.5}, where pressure maxima
are observed at the jet center and in the interaction of the growing
wave with the ambient.

\subsection{Fully non-linear regime}

\begin{figure}[!t]
\centerline{
\psfig{file=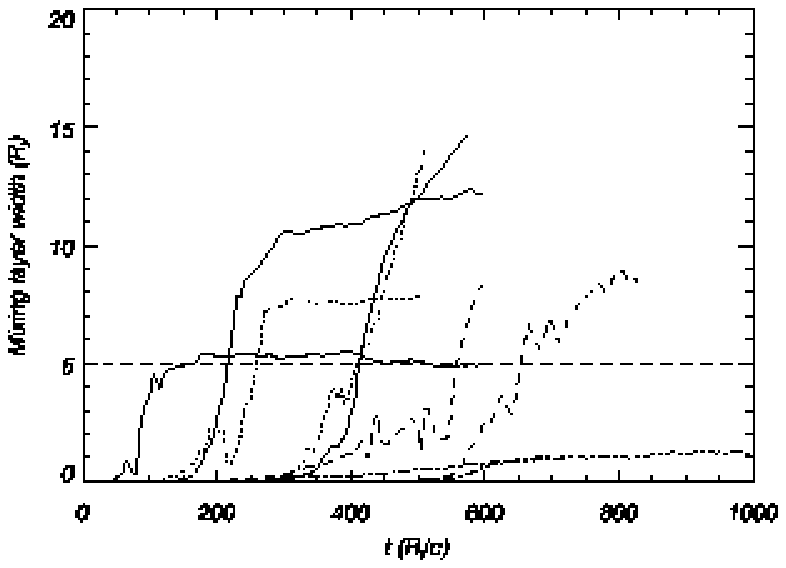,width=0.5 \textwidth,angle=0,clip=}}

\caption{Evolution of the mean width of the jet/ambient mixing layer
(for tracer values between 0.05 and 0.95) with time for all
simulations. Lines represent the same models as in
Fig. \ref{fig:f2}. A value of $5\,R_j$ for the width of the mixing
layer (horizontal dashed line) serves to classify the evolution of the
different models.}
\label{fig:f1}

\end{figure}

  In Paper~II, the non-linear evolution of the instability in the
different models was characterized through the processes of
jet/ambient mixing and momentum transfer. In Fig.~\ref{fig:f1} we show
the width of the mixing layer as a function of time for all the
models. The times at which mixing starts in the different models
$t_{\rm mix}$ appear listed in Table~\ref{tab:phases}. In all cases
these times are around $t_{\rm sat}$. Generally, models developing
wider shear layers are also more mixed; i.e., the amount of mass in
zones with jet mass fraction strictly different from 0 and 1, is
higher. We observe that those models with larger values of
$\Delta_{peak}$ (lower Lorentz factor and colder jets) develop wider
layers ($>5R_j$) soon after saturation due to turbulent mixing
induced by the shock, while those models where resonant modes appear
do not show strong mixing with the ambient. Models \textbf{B10} and
\textbf{D10} undergo a mixing process, though slower than the former.
%
%
\begin{figure}
\centerline{
\psfig{file=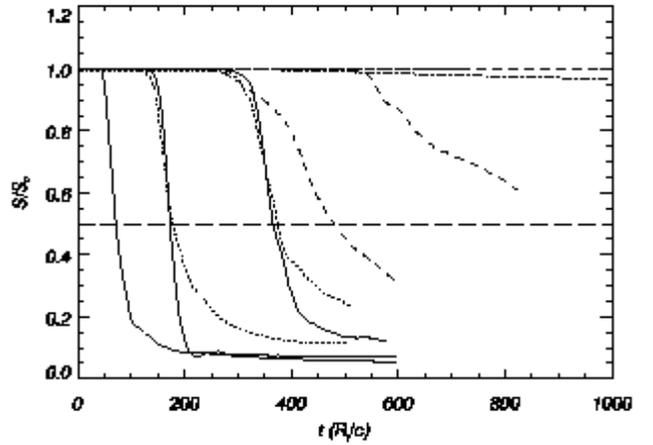,width=0.5 \textwidth,angle=0,clip=}}

\caption{Evolution of the normalized total longitudinal momentum in
the jet as a function of time. Lines represent the same models as in
Fig. \ref{fig:f2}. The long-dashed horizontal line
identifies those models transferring more than 50 \% of the initial jet
momentum to the ambient.}
\label{fig:f3}

\end{figure}
%

\begin{figure}
\centerline{
\psfig{file=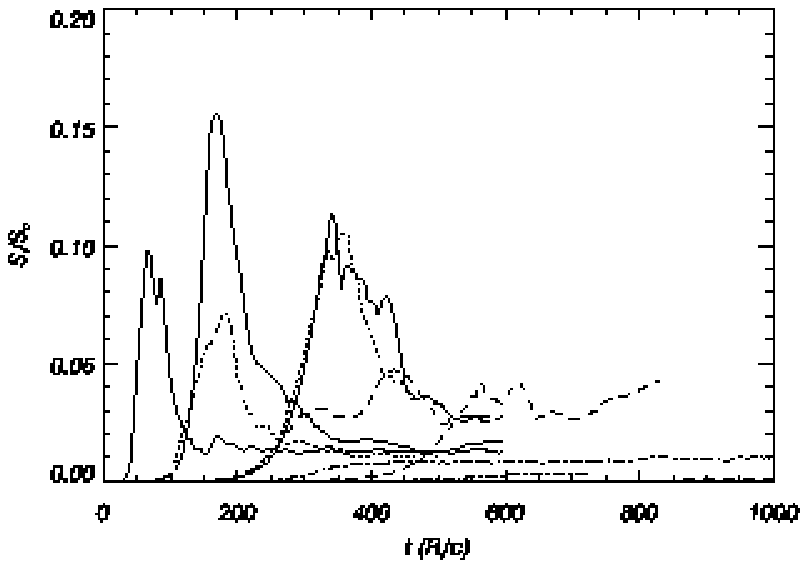,width=0.5 \textwidth,angle=0,clip=}}

\caption{Evolution of the normalized total transversal momentum in the
jet as a function of time for all the simulations. Lines represent the
same models as in Fig.~\ref{fig:f2}.}
\label{fig:f4}

\end{figure}

  Figure~\ref{fig:f3} shows the fraction of initial axial momentum
kept by the jet as a function of time. Axial momentum is lost first
through sound waves forming the linear perturbations and
second, but more important, through shocks themselves and by
subsequent mixing, which implies sharing of momentum with the ambient
medium. Correlation with Fig.~\ref{fig:f1} is remarkable. Models
developing wide mixing layers coincide with those losing more than
$50\%$ of their initial axial momentum just after saturation; models
\textbf{B10} and \textbf{D10} share their momentum with the ambient
medium continuously in the non-linear regime; and models where
resonant modes dominate saturation keep almost all their initial
momentum by the end of the simulations. Results derived from
Fig.~\ref{fig:f3} are corroborated by Fig.~\ref{fig:f4}. In the latter
we plot the total transversal momentum in the jets normalized to the
corresponding longitudinal momentum. Transversal momentum in the jet
(initially zero) is generated through turbulent motions and continuous
conversion of kinetic into internal energy. The value of the
normalized transversal momentum at a given time is an indication of
how far from equilibrium the jet is. We observe that colder and lower
Lorentz factor models present strong peaks at $t_{\rm sat}$,
coincident with the triggering of the shock and the sudden transfer of
longitudinal momentum seen in the previous plot: Those models where
resonant modes appear barely generate any transversal momentum, and
models \textbf{B10} and \textbf{D10} do not present strong peaks at
saturation but display a steady transmission of the transversal momentum
through the non-linear regime (see Fig.~\ref{fig:f4}), implying continuous
loss of energy.
%
%

\begin{figure*}
\centerline{
\psfig{file=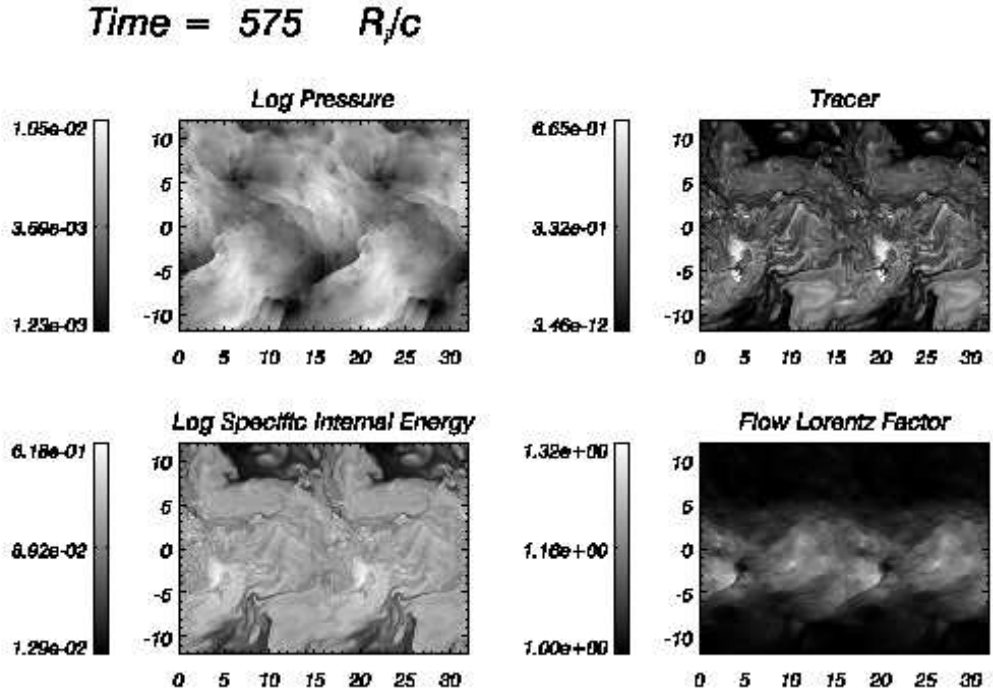,width=0.8\textwidth,angle=0,clip=}}

\caption{Snapshot at the last frame of the simulation of logarithmic
maps of pressure, jet mass fraction and specific internal energy and
non-logarithmic Lorentz factor for model \textbf{A2.5}.}
\label{fig:f6}

\end{figure*}

\begin{figure*}
\centerline{
\psfig{file=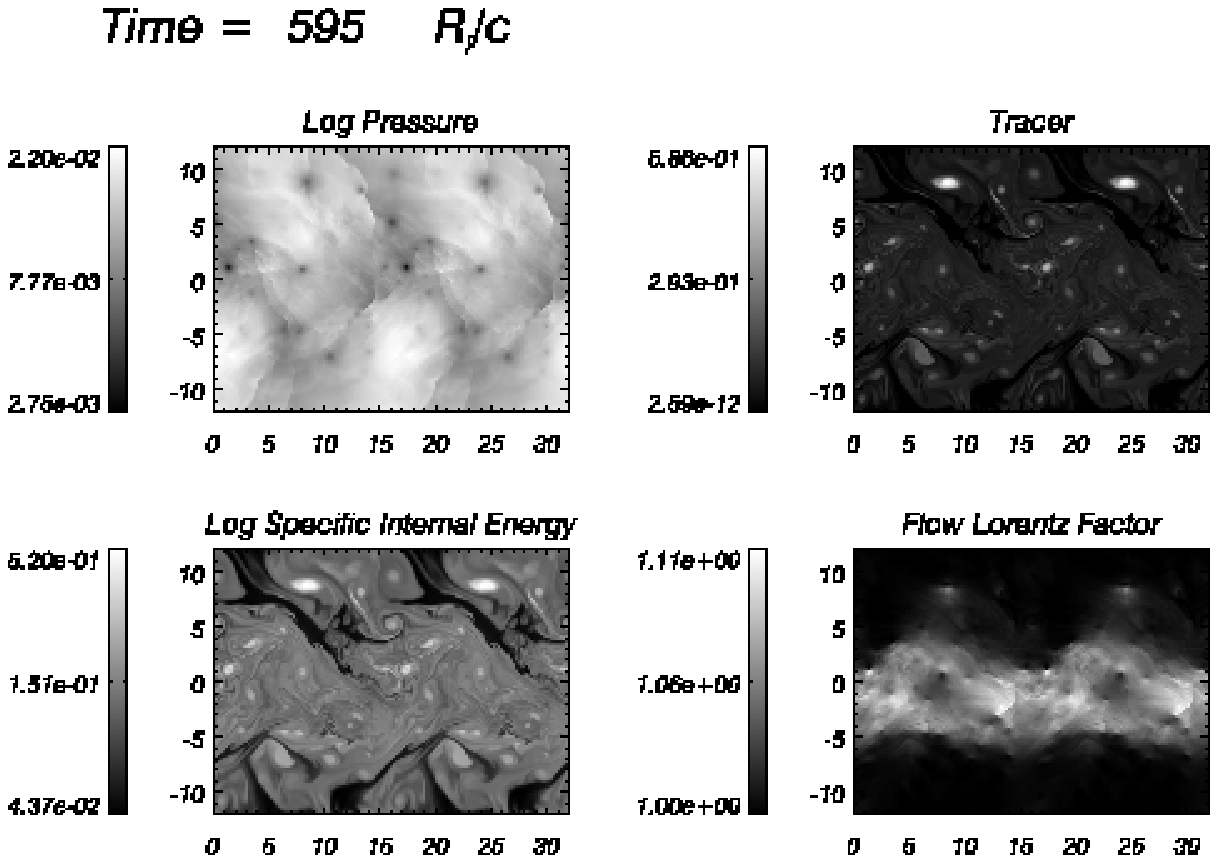,width=0.8\textwidth,angle=0,clip=}}

\caption{Same as Fig. \ref{fig:f6} for model \textbf{B2.5}.}
\label{fig:f8}

\end{figure*}

\begin{figure*}
\centerline{
\psfig{file=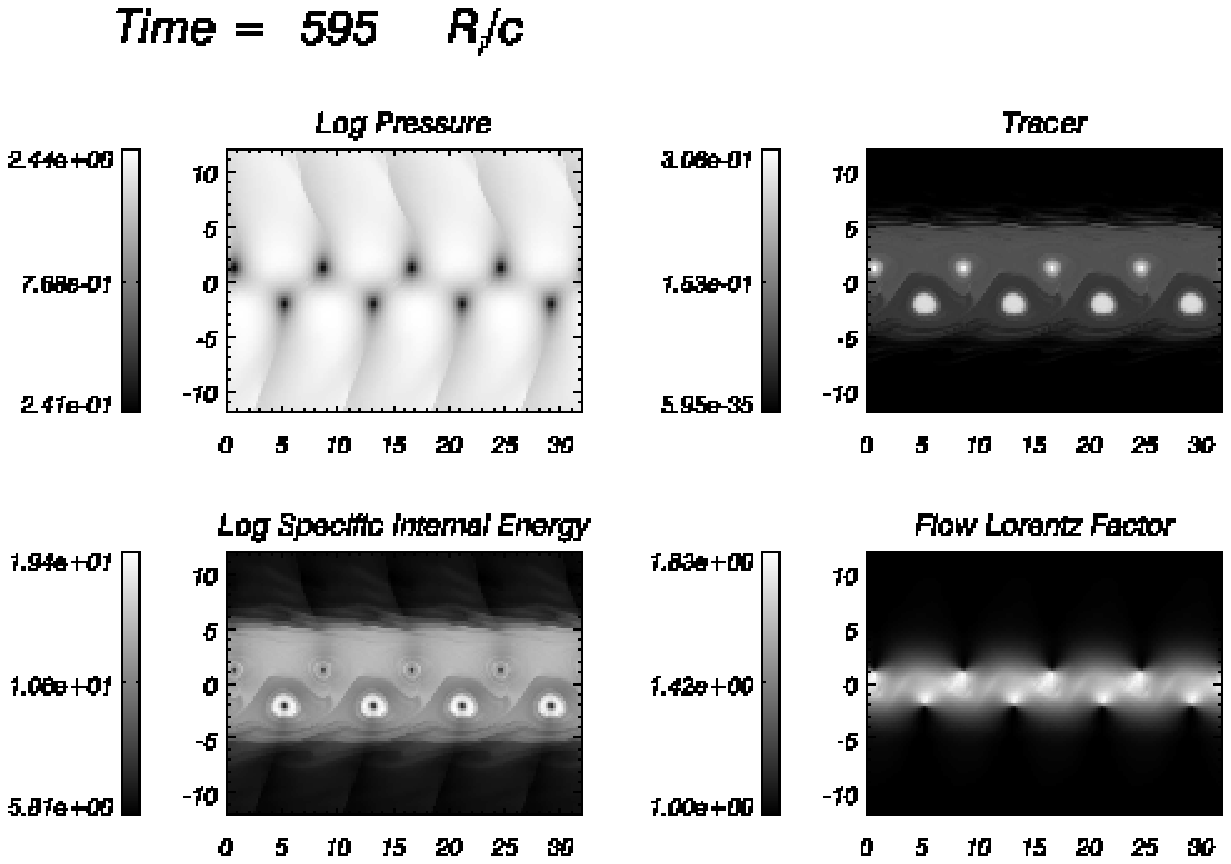,width=0.8\textwidth,angle=0,clip=}}

\caption{Same as Fig. \ref{fig:f6} for model \textbf{D2.5}.}
\label{fig:f12}

\end{figure*}

\begin{figure*}
\centerline{
\psfig{file=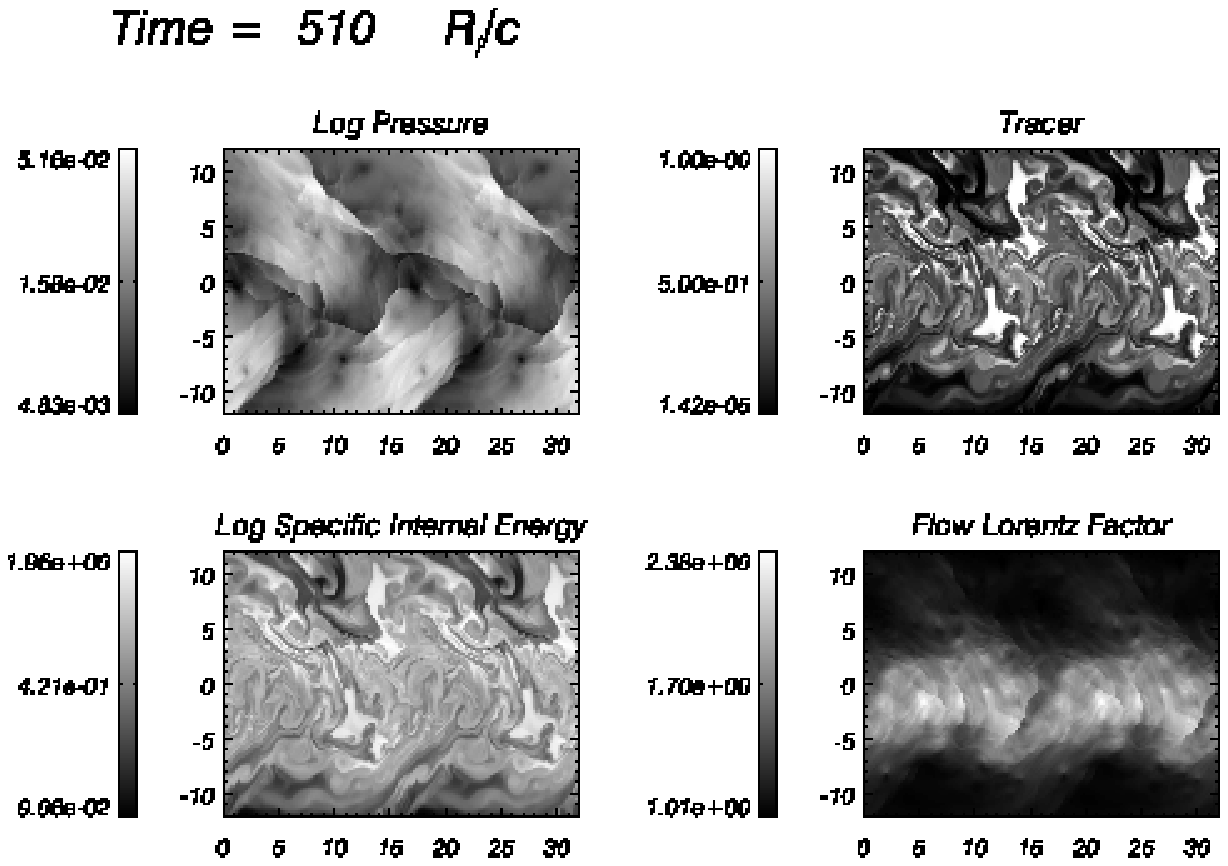,width=0.8\textwidth,angle=0,clip=}}

\caption{Same as Fig. \ref{fig:f6} for model \textbf{B05}.}
\label{fig:f9}

\end{figure*}

\begin{figure*}
\centerline{
\psfig{file=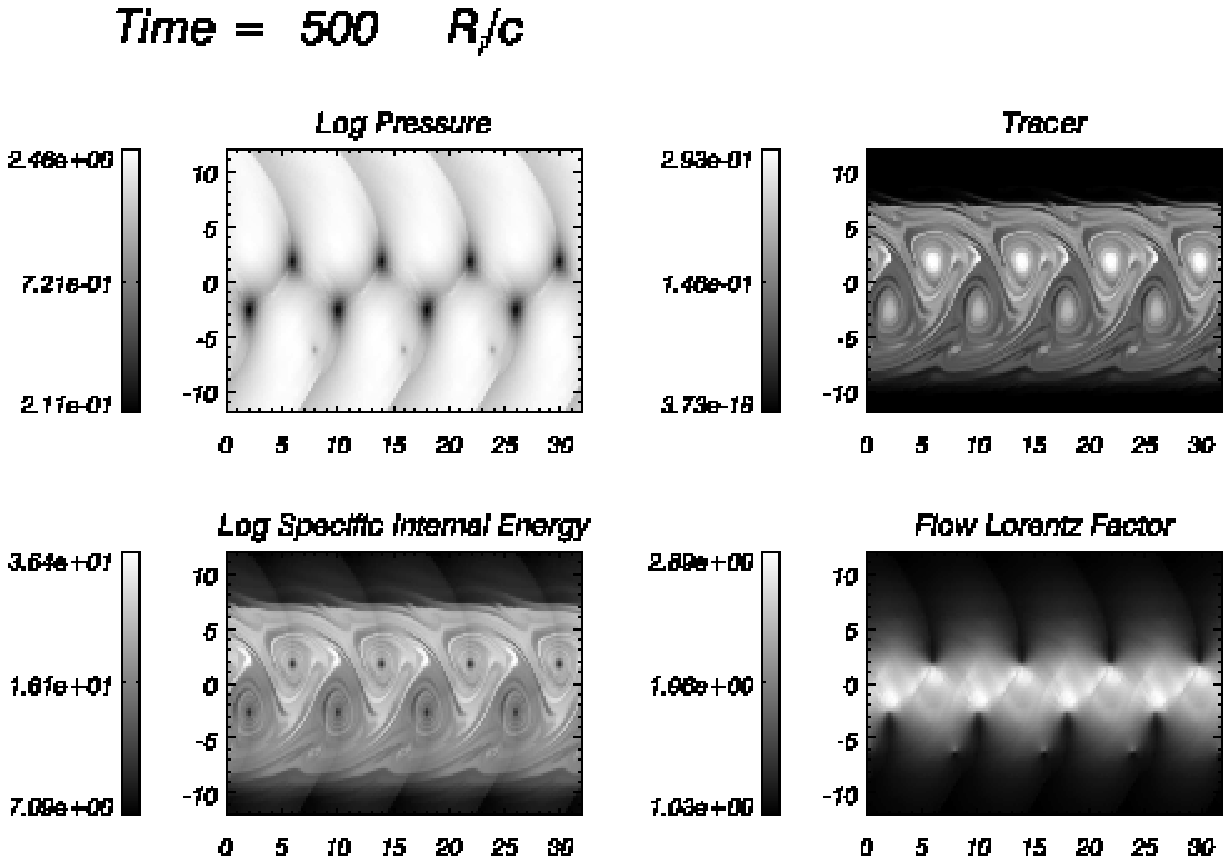,width=0.8\textwidth,angle=0,clip=}}

\caption{Same as Fig. \ref{fig:f6} for model \textbf{D05}.}
\label{fig:f13}

\end{figure*}

\begin{figure*}
\centerline{
\psfig{file=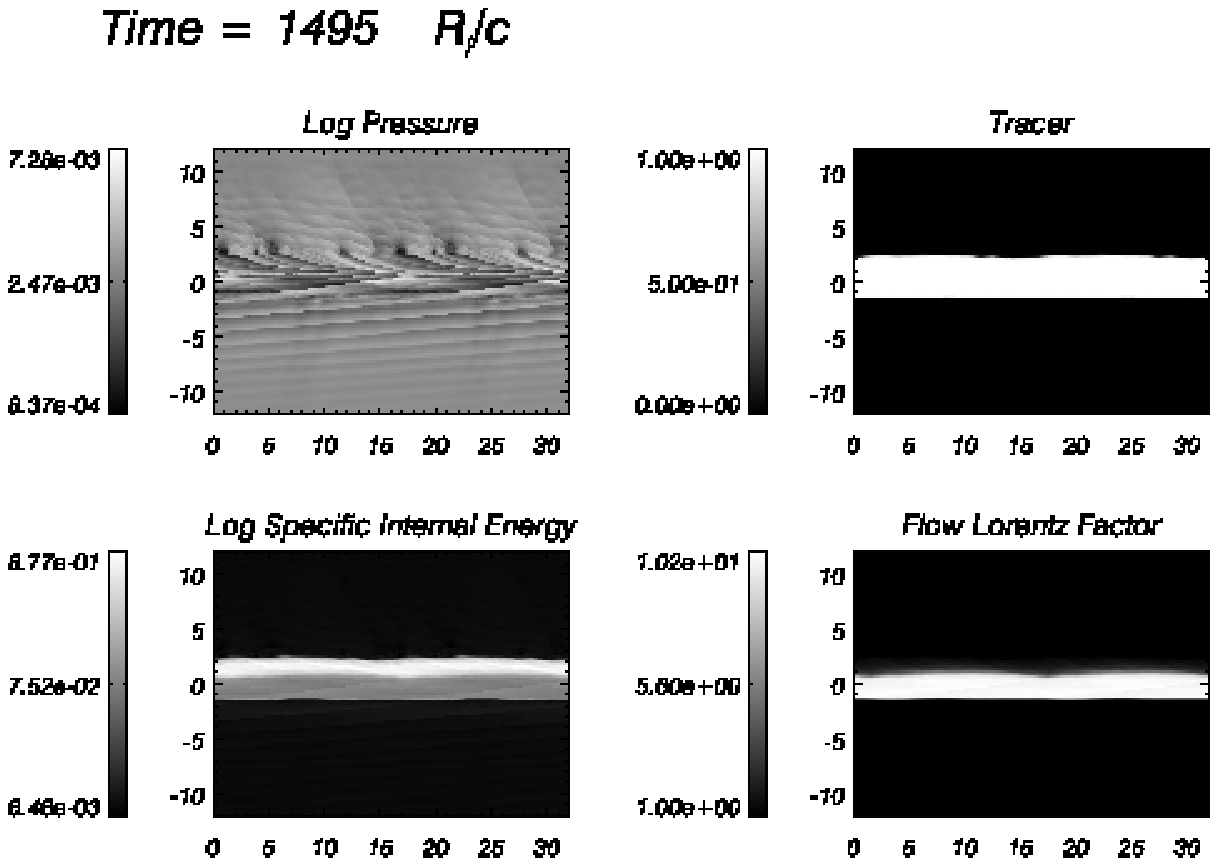,width=0.8\textwidth,angle=0,clip=}}

\caption{Same as Fig. \ref{fig:f6} for model \textbf{A10}.}
\label{fig:f7}

\end{figure*}

\begin{figure*}
\centerline{
\psfig{file=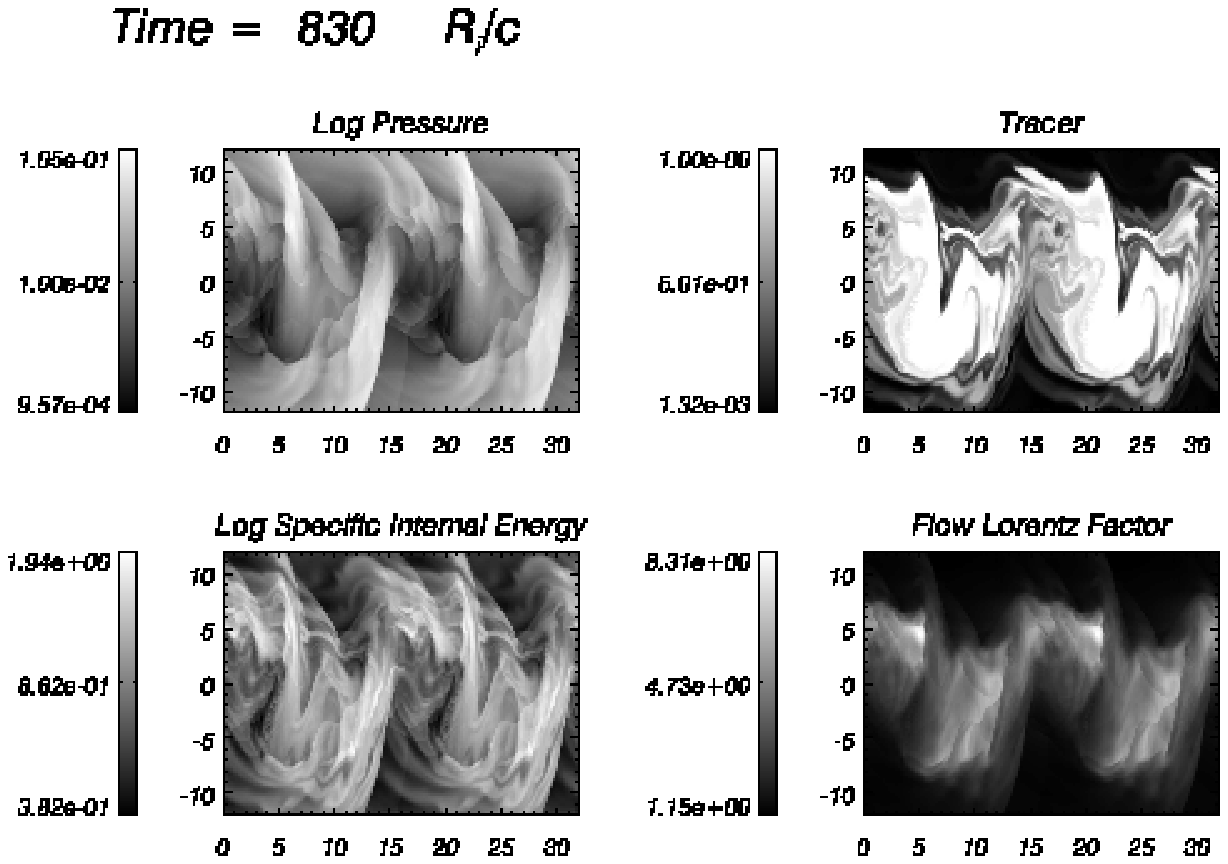,width=0.8\textwidth,angle=0,clip=}}

\caption{Same as Fig. \ref{fig:f6} for model \textbf{B10}.}
\label{fig:f10}

\end{figure*}

\begin{figure*}
\centerline{
\psfig{file=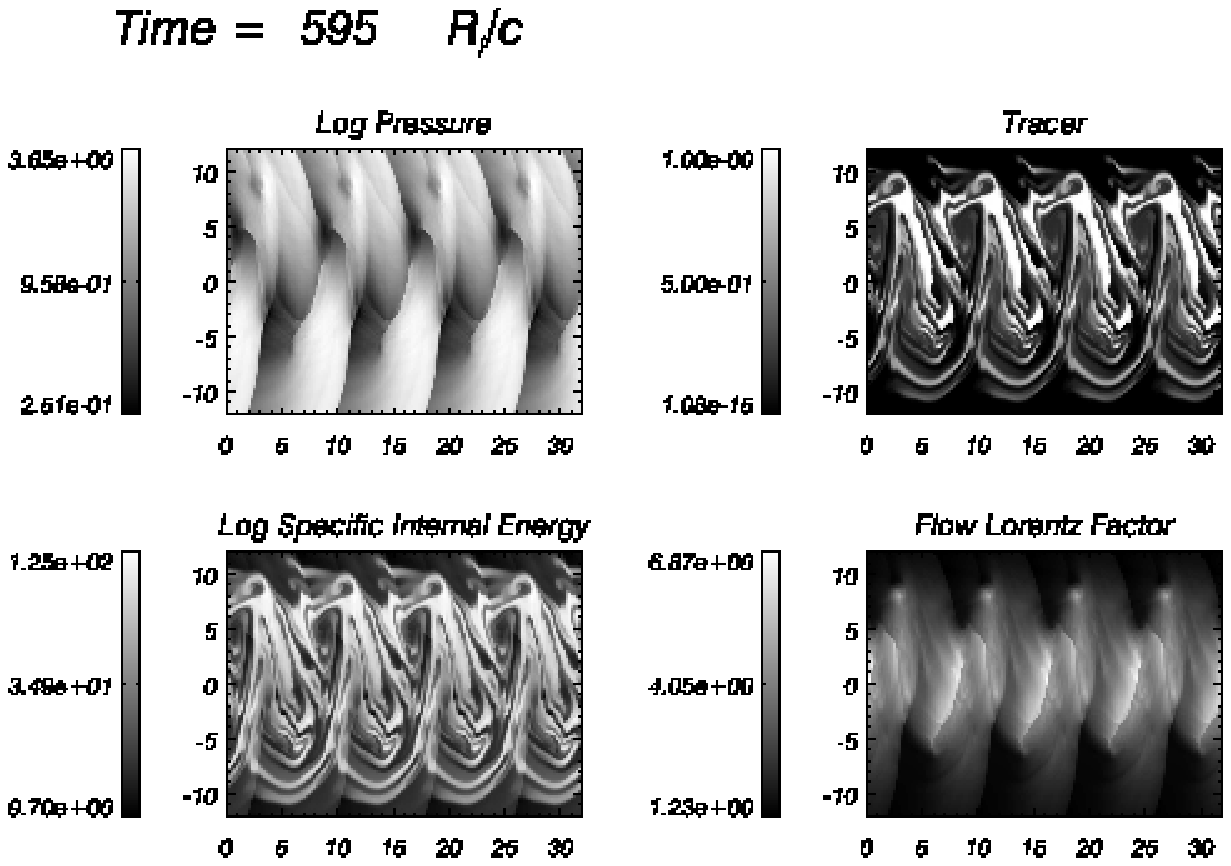,width=0.8\textwidth,angle=0,clip=}}

\caption{Same as Fig. \ref{fig:f6} for model \textbf{D10}.}
\label{fig:f14}

\end{figure*}

\begin{figure*}
\centerline{
\psfig{file=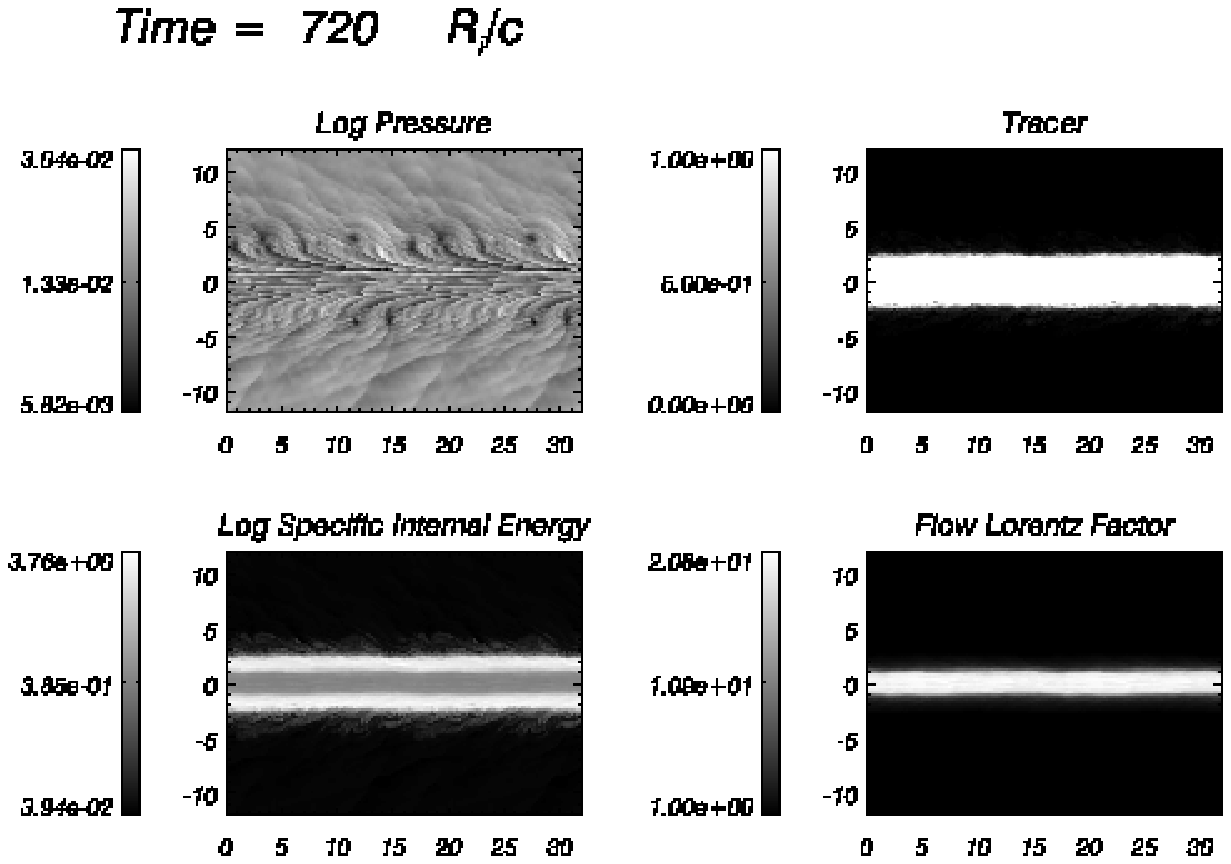,width=0.8\textwidth,angle=0,clip=}}

\caption{Same as Fig. \ref{fig:f6} for model \textbf{B20}.}
\label{fig:f11}

\end{figure*}

\begin{figure*}
\centerline{
\psfig{file=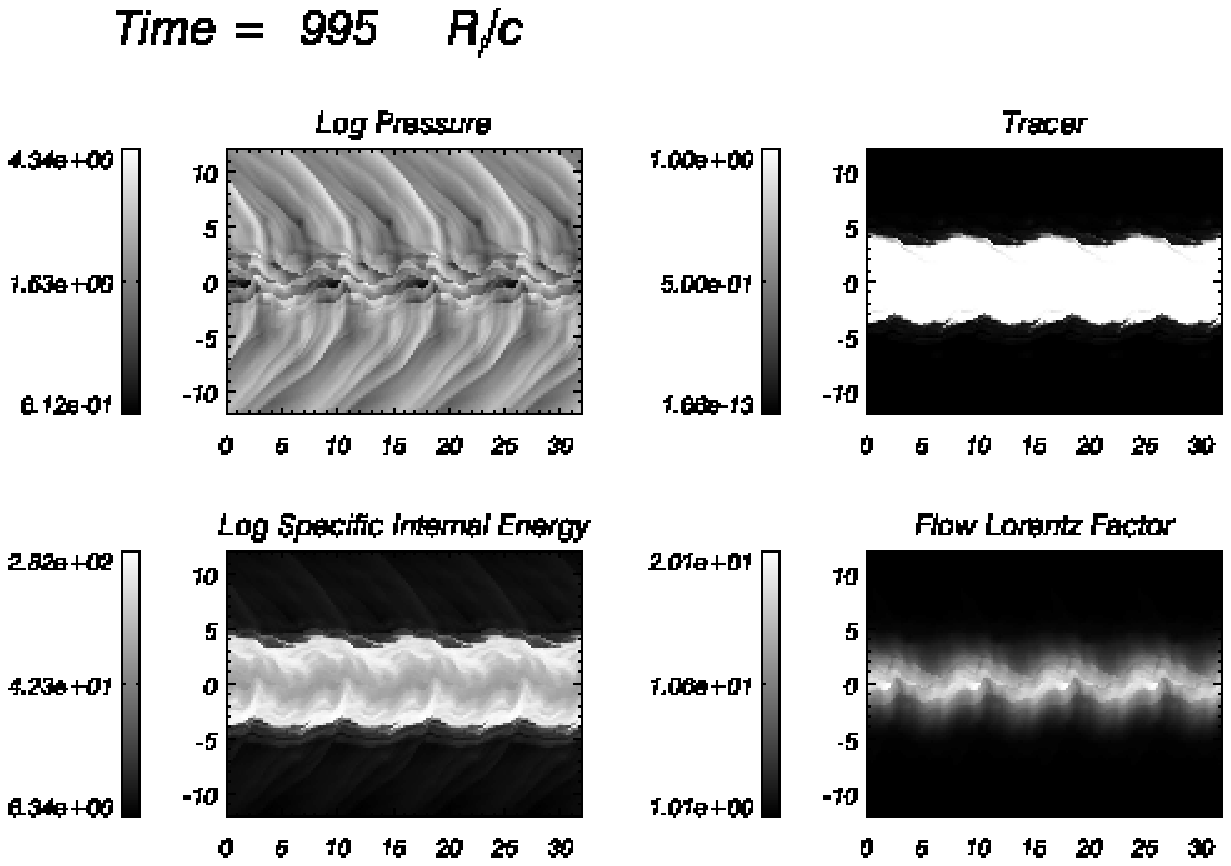,width=0.8\textwidth,angle=0,clip=}}

\caption{Same as Fig. \ref{fig:f6} for model \textbf{D20}.}
\label{fig:f15}

\end{figure*}

  Panels showing several physical quantities for all models at the end
of simulations are presented in Figs.~\ref{fig:f6}-\ref{fig:f15}.
Colder and slower models (\textbf{A2.5}, \textbf{B2.5}, and
\textbf{B05}) show turbulent mixing in a wide region and are barely
relativistic by the end of the simulations. Models \textbf{D2.5} and
\textbf{D05} have mixed deeply (the jet mass fraction is less than one
everywhere) but keep larger Lorentz factors. Moreover, these models
seem to have stopped the widening process of the mixing layer as it is
deduced from the flattening of the mixing layer width as a function of
time in Fig.~\ref{fig:f1}. Models \textbf{B10} and \textbf{D10} are
also undergoing turbulent mixing. From Figs.~\ref{fig:f1} and
\ref{fig:f3}, it is deduced that \textbf{B10} and \textbf{D10} are
still mixing and transferring momentum by the end of simulations.
These models will eventually lose a large amount of their initial
longitudinal momentum, thereby becoming colder and denser due to mass
entrainment from the ambient medium. Finally, models \textbf{A10},
\textbf{B20} and \textbf{D20} present a fast core $\sim 1\,R_j$ wide
with rich internal structure as a consequence of the resonant modes
(see subsection on the linear regime), which is surrounded by a hot
and slow shear layer that extends up to $\sim 2\,R_j$ in models
\textbf{A10} and \textbf{B20} or $\sim 4\,R_j$ in model
\textbf{D20}. Let us point out that model \textbf{A10}
(Fig.~\ref{fig:f7}) displays a highly asymmetric structure, resulting
from the development of resonant modes only on the upper
interface. This is a consequence of the combination of symmetric and
antisymmetric modes, and probably of nonlinear interactions between
resonant modes, which result in destructive interference on one side
of the jet and constructive interference on the other.

%
%

%
%

%
%

%
%

%
%

%
%

%
%

%
%

%
%

%
%

\section{Discussion}
\label{sect:disc}

\subsection{Non-linear stability}

  Simulations presented in Papers I and II, performed for the most
unstable first reflection mode of the corresponding models, confirmed
the general trends resulting from the linear stability analysis: the
faster (larger Lorentz factor) and colder jets have smaller growth
rates in the linear regime. In Paper~II, the non-linear evolution of
the instability in the different models was characterized through the
processes of jet/ambient mixing and momentum transfer. The models were
then classified into four classes (I to IV) according to the
particular nature of these processes in each of the models. Class I
models (corresponding to cold and slow jets) were deeply mixed and
mass-loaded by the end of the simulation. Models in Class II (hot and
fast jets) were slowly mixed in the nonlinear regime, progressively
losing their longitudinal momentum. Models in Class III (hot and slow
jets) have properties between Classes I and II. Finally, Class IV
(containing cold/warm and fast models) appeared as the most stable in
the nonlinear regime. Shear layers formed in all the models as a
result of the non-linear evolution. Models in Classes I/II developed
broad shear layers and appeared totally mixed, cooled, and slowed
down. In contrast, models in Classes III/IV have an inner core
surrounded by thinner layers and keep a larger amount of their initial
longitudinal momentum. We performed a number of additional simulations
keeping the properties of the ambient medium fixed and changing the
rest-mass density of the jet and the Lorentz factor. Results confirmed
that these models behave like previous simulations, and are naturally
placed in the classification already defined.
%
\begin{figure*}
\centerline{
\psfig{file=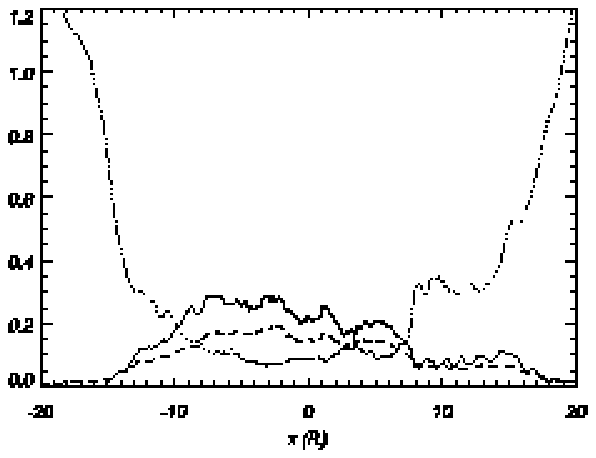,width=0.4
\textwidth,angle=0,clip=} \quad \psfig{file=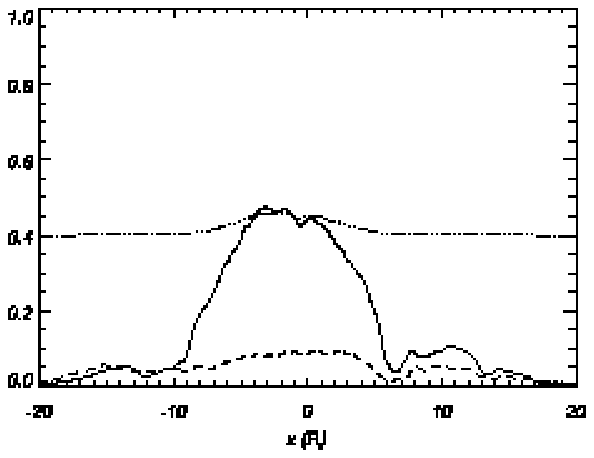,width=0.4
\textwidth,angle=0,clip=}}

\centerline{
\psfig{file=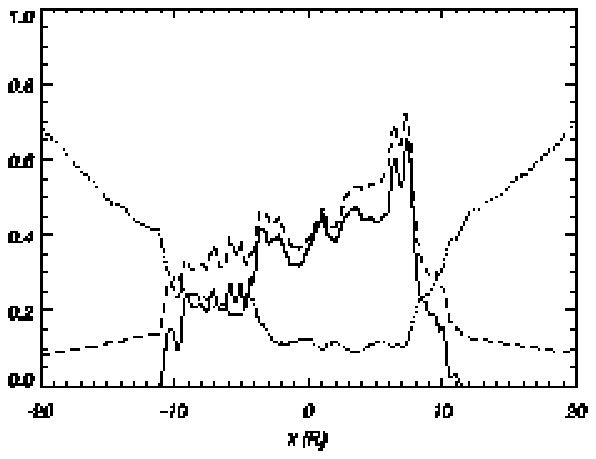,width=0.4
\textwidth,angle=0,clip=} \quad \psfig{file=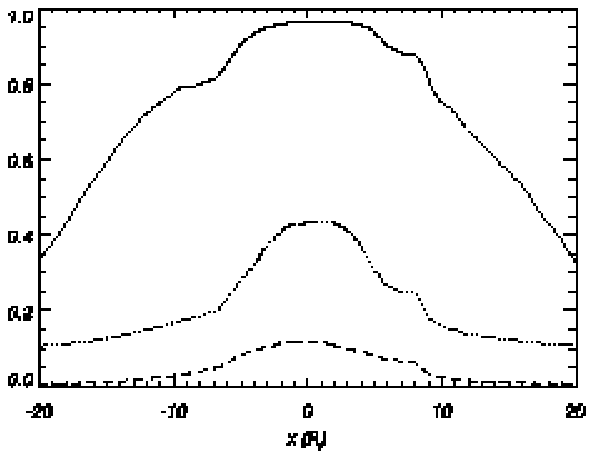,width=0.4
\textwidth,angle=0,clip=}}

\centerline{
\psfig{file=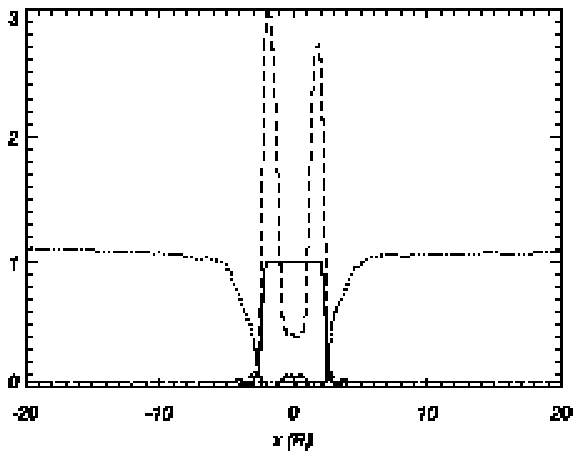,width=0.4
\textwidth,angle=0,clip=} \quad \psfig{file=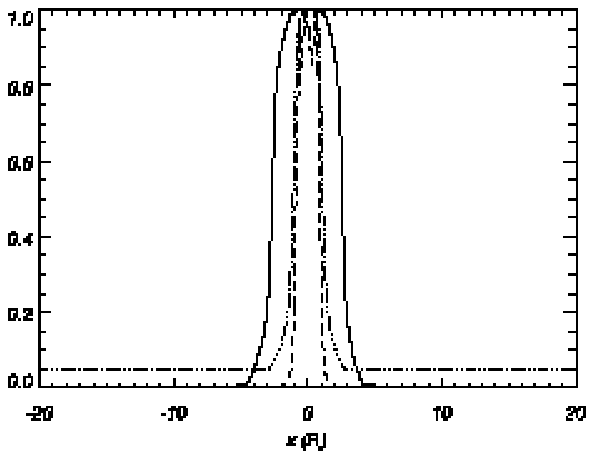,width=0.4
\textwidth,angle=0,clip=}}

\caption{Averaged transversal structure in the final state of the jets
corresponding to models \textbf{A2.5} (upper panels), \textbf{D10}
(middle), and \textbf{B20} (bottom). Left panels (thermodynamical
quantities): solid line, tracer; dotted line, rest mass density;
dashed line, specific internal energy. Right panels (dynamical
quantities): solid line, longitudinal velocity; dotted line, lorentz
factor normalized to the initial value in the jet; dashed line,
longitudinal momentum normalized to the initial value in the
jet. Specific internal energy for model \textbf{D10} was divided
by $100$ to fit in the scale.}
\label{fig:profs}

\end{figure*}
%

  The stability classes considered in Paper~II were defined according to
the jet response to single modes. In this paper we revisit this
classification scheme in the light of the present results based on
more general perturbations. From the analysis of Figs.~\ref{fig:f1},
\ref{fig:f3}, and \ref{fig:f4}, we classified jets depending on
their nonlinear behaviour in three different groups:

\begin{itemize}

  \item Unstable 1 (UST1) models: jets which are disrupted after a
strong shock is formed after the linear regime, enhancing turbulent
mixing with ambient medium. It includes models \textbf{A2.5},
\textbf{B2.5}, \textbf{D2.5}, \textbf{B05}, and \textbf{D05}, i.e.,
lower Lorentz factor jets. The mixing layer width becomes larger than
$5\,R_j$ (Fig. \ref{fig:f1}), and they share more than $50\%$ of their
initial momentum with the ambient medium (Fig. \ref{fig:f3}).

  \item Unstable 2 (UST2) models: jets which are disrupted in the
non-linear phase by a continuous process of momentum transfer to the
external medium, like \textbf{B10} and \textbf{D10}. This is observed
in Fig. \ref{fig:f4} as a non-decreasing transversal momentum in the
nonlinear regime. These models eventually end up sharing a large
fraction of initial momentum and developing a wide mixing layer.

  \item Stable (ST): jets which develop resonant modes and remain
collimated for long time, \textbf{A10}, \textbf{B20}, and
\textbf{D20}. These models have a thin mixing layer and share a very
small fraction of their axial momentum with the ambient medium. They
expand, but remain collimated.

\end{itemize}

  In the course of their evolution, the jets develop a rich transversal
structure in all the physical variables. This structure is different
depending on the non-linear evolution of the jets.
Figure~\ref{fig:profs} displays the transversal profiles of relevant
physical quantities averaged along the jet at the end of the
simulations for model \textbf{A2.5}, representative of models in
UST1, \textbf{D10} of UST2, and \textbf{B20} of ST.

  Model \textbf{A2.5} shows a totally mixed, mass-loaded flow with
averaged maximum speed $0.4c$, i.e., barely relativistic, as these
jets are efficiently slowed down by mass entrainment after the
disruption. The mass loading is inferred from the low values in the
tracer profile ($f<0.3$), which imply a large fraction of ambient medium
material inside the jet. The efficient conversion of kinetic energy
into internal energy enhanced by the shock triggered in the early
post-linear phase causes the jet to increase its specific internal
energy.

  UST2 jets undergo a slower process of mixing, so they still keep a
larger fraction of axial velocity and Lorentz factor by the end of the
simulation, even though they appear to be totally mixed ($f<0.7$
everywhere). However, as we have mentioned in the previous section,
the mixing and slowing process is still going on in \textbf{B10} and
\textbf{D10}, so it is clear that if the simulation had continued, the
longitudinal velocity and Lorentz factor values would be smaller than
those found. We also observe that the more mass-loaded parts of the
jet (i.e., the region with $-10\,R_j<x<0$) are consistently colder.

  Finally, the jet in model \textbf{B20} remains very thin. The
velocity profile of the model has widened by $2-3\,R_j$ by the end of the
simulation, coinciding with the generation of a hot shear layer. This
layer is seen in the figure as an overheated and underdense
($\rho<0.1$) region shielding the unmixed core ($f=1.$), which keeps
almost all its initial axial momentum and Lorentz factor. The core has
a rich internal structure (see the pressure panel in
Fig.~\ref{fig:f11}) that also manifests in the spiky structure of the
shows longitudinal momentum profile.

  A comparison of the present non-linear evolution classification
scheme and that of Paper II (classes I-IV) allows us to conclude that,
in general, models in Classes I and III fall into UST1, whereas models
in Class II corresponds to UST2 and those in Class IV to ST.  The
reason models \textbf{D2.5} and \textbf{D05} (belonging formerly to
Class III) move to UST1 may be the inclusion of longer wavelength
perturbations, along with the antisymmetric modes, which are more
disruptive than the symmetric first reflection mode used in the
previous work, and the lack of axial resolution in the latter, as
discussed in the Appendix of Paper II. This can be seen by comparing
structures and evolution of model \textbf{D05} here and in Paper II,
in particular the evolution of the mixing and momentum transfer.

  Regarding UST2 here compared to former class II, \textbf{B10} and
\textbf{D10} undergo a very similar slow process of momentum transfer
to the external medium to that observed for D10 and D20 in Paper II,
although their temperatures are very different and the shock in
\textbf{B10} is much stronger than in \textbf{D10} (see
Table~\ref{tab:phases}). The reason for this slow momentum exchange
may be the same as proposed in Paper II for models D10 and D20,
i.e., a continuous conversion of kinetic into internal energy due to
the large initial Lorentz factor, which acts as a source of transversal
momentum favoring the process of mixing and mass-loading.
Model \textbf{B10} changes from Class I in Paper II to UST2 here, meaning
that disruption occurs by slow mixing in the new simulation, compared
to sudden disruption in the previous one.

  Models in Class IV were characterized by a rich internal-structure
jet preserving a large fraction of initial momentum and Lorentz
factor. ST models share these features, but now we are able to
clearly associate them to with the growth of resonant modes, which
could be the reason for the breaking of the linear slope in model
C20 in Paper ~I (see Fig.~2 there). Steepening of short wavelength
perturbations at the shear layer generates small shocks which
favor local mixing and an efficient conversion of kinetic into
internal energy. As a result of this process, the shear layer
heats up and the jet expands to form a hot and underdense layer
around the jet core (see Fig.~\ref{fig:profs}). It is remarkable
that model \textbf{A10} is largely asymmetric by the end of the
simulation (see Fig.~\ref{fig:f7}). This is a consequence of the
resonant modes only growing on one side of the jet during the
linear regime, and it is understood on the basis of asymmetry
resulting from mixture of symmetric and antisymmetric modes. This
effect, though much less evident, is also observed in model
\textbf{B20}. Finally, model \textbf{D20} has moved from class II
in Paper II to ST here, clearly due to the appearance of resonant
modes. This fact allows us to conclude that the fate of ST models
would be exactly the same as those in UST2, if it were not for the
growth of resonant modes; hence, their importance in the long term
stability of these jets.

%
\begin{figure}
\centerline{
\psfig{file=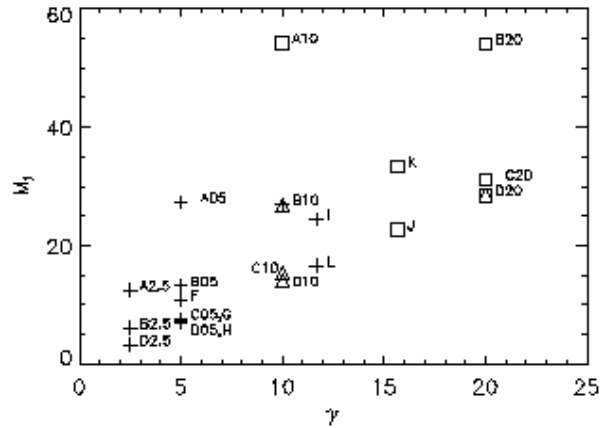,width=0.5
\textwidth,angle=0,clip=}}

\caption{Relativistic internal jet Mach number ($M_j$) versus jet
Lorentz factor ($\gamma$) of the simulated models here and in Papers I
and II. Symbols represent different non-linear behaviors: crosses
stand for UST1 disrupted jets (low relativistic Mach number and low
Lorentz factor); triangles for UST2 jets (moderately fast and
supersonic), and squares for ST jets (highly supersonic and fast
jets). Models with two different symbols are those with a different
evolution in simulations presented here and those from Papers I and II
(see text).}
\label{stabplane}

\end{figure}
%

  We classified jets depending on their nonlinear behaviour in
three different groups, which are clearly separated in a relativistic
internal jet Mach number vs. jet Lorentz factor plot
(Fig.~\ref{stabplane}). In this plot, we also include models from
Paper II, in order to show the general character of our results and to
note that this division of the stability properties of jets is more
accurate than in Paper II, with the jet-to-ambient enthalpy ratio
instead of the relativistic Mach number. A clear correlation between
the two plotted parameters and the non-linear stability properties of
the jets is observed. Models \textbf{B10} and \textbf{D20} are
represented with two different symbols to show the change of nonlinear
behaviour from Paper II. These are placed in transition regions of the
plot, either in Lorentz factor (\textbf{B10}) or in relativistic Mach
number (\textbf{D20}). This fact could explain differences in the
non-linear behaviour as caused by changes in the initial jet profiles,
what is quite evident in the case of \textbf{D20}, for resonant modes
appear due to the presence of the shear layer. As in the previous
discussion, we have given the same symbols (crosses) as for UST1 jets
here to models in Class III of Paper II, as we do not consider that
they have different non-linear behaviour in both
simulations. Figure~\ref{stabplane} can be considered as the
relativistic counterpart of the $M-\nu$ (Mach number-density ratio)
diagram in Bodo et al. (1994); note that the density ratio,
$\nu=\rho_a/\rho_j$, is inverted with respect to ours. In our case
the Mach number is relativistic; and the density ratio, which stands
for the inertia of the flow, is replaced by the Lorentz factor
here, as relativistic momentum is $\propto \gamma^2$, so it dominates
the inertia of relativistic jets. Our conclusions are similar to
theirs, for denser (higher Lorentz factor) and highly supersonic jets
(high relativistic Mach number) are the stablest. However, in our
case, we found a higher degree of stability due to the growth of
resonant, stabilizing modes.

This result agrees with the conclusion of Hardee (2000),
where linear stability arguments show that distortions
induced by instabilities are smaller for higher Lorentz factor flows,
although they were not associated to the shear resonances reported by
us.

Finally, simulations discussed in Appendices A (single
antisymmetric mode in planar geometry) and B (single symmetric
mode in cylindrical geometry) have confirmed general trends of the
present clasification scheme, generalizing our results.

\subsection{Astrophysical implications}

  One of the current open problems in extragalactic jet research is to
understand the morphological dichotomy between FRI and FRII jets.
Several possible explanations have been proposed which mainly fall in
one of these two general possibilities: either (i) FRI and FRII
sources are intrinsically the same, and the morphology and jet
evolution depend mainly on the ISM in which they are embedded in the
first kiloparsecs, or (ii) they depend on intrinsic differences
stemming from the jet formation process (black hole rotation,
Blandford 1994; accretion rate, Baum et al. 1995; black hole mass,
Ghisellini \& Celotti 2001; the so-called magnetic switch, Meier et
al. 1998), or (iii) a combination of both (e.g., Snellen \& Best
2003). Of course, all the mechanisms could come into play with
differing effects and significance depending on the source.

  Leaving the basis of the morphological dichotomy aside, current
models (Laing \& Bridle 2002a,b and references therein) interpret FRI
morphologies as the result of a smooth deceleration from relativistic
($\gamma \leq 3$, Pearson 1996) to non-relativistic transonic speeds
($\sim 0.1\,c$) on kpc scales. On the contrary, flux asymmetries
between jets and counter-jets in the most powerful radio galaxies and
quasars indicate that relativistic motion ($\gamma\sim 2-10$) extends
up to kpc scales in these sources, although with smaller values of the
overall bulk speeds ($\gamma \sim 2-4$, Bridle et al. 1994). Current
models for high energy emission from powerful jets at kpc scales
(e.g., Celotti et al. 2001) offer additional support to the hypothesis
of relativistic bulk speeds on these scales.

  The results concerning the long-term evolution of relativistic jets
presented in this paper and summarized in Fig.~\ref{fig:profs}
confirm that slower and smaller Mach number jets (UST1) are entrained
by ambient material and slowed down to $v<0.5\,c$ after becoming
overpressured (due to conversion of kinetic into internal energy) and
being disrupted by nonlinear instabilities effects which cause flaring
and rapid expansion of the mixing flow. UST2 jets undergo a smooth
slowing down; and though by the end of the simulation jet velocity is
$\sim 0.9\,c$, this process is continuous, and eventual loss of
velocity to mildly relativistic values is to be expected. Finally, ST
jets keep their initial highly relativistic velocities, and their
steadiness by the end of simulations makes them firm candidates for
remaining collimated over long distances. Hence our results would
point to a high Lorentz factor, highly supersonic jets as forming FRII
Class, whereas FRI jets would be found in the opposite corner of the
diagram (slow, small Mach number jets). The validity of our results
extends to models with different jet-to-ambient-density ratios and
specific internal energies as seen in Paper~II.

  Our conclusions point to an important contribution by intrinsic
properties of the source to the morphological dichotomy. Nevertheless,
the importance of the ambient medium cannot be ruled out on the basis
of our simulations, since we consider an infinite jet in pressure
equilibrium flowing in an already open funnel and surrounded by a
homogeneous ambient medium. Thus our approach does not take into
account the consequences of the interaction of the jet with the
ambient in order to penetrate it or the effects of a spatially varying
atmosphere. Simulations following the spatial approach (perturbations
grow with distance) for jets propagating in different ISM profiles and
using a more realistic microphysics (allowing for a local mixture of
electron, positron, and proton Boltzmann gases) will be performed in
order to clarify these points.

  As dicussed in the introduction of this paper, there are plenty of
arguments indicating the existence of transversal structure in
extragalactic jets at all scales. In the simulations presented here,
the initial states were defined with a continuous transition layer of
thickness $\approx 0.2 R_j$. As discussed in the previous paragraphs,
this shear layer has played an important role in the long-term
stability of the jet flow. Besides this, thicker shear layers have
been generated in the course of the non-linear evolution. Relatively
thin ($\approx 2 R_j$), hot shear layers are found in present ST
models (the physically meaningful counterparts of the layers found in
the three-dimensional, low-resolution simulations of Aloy et al. 1999,
2000), which could explain several observational trends in powerful
jets at both parsec and kiloparsec scales (see Aloy et al. 2000 and
references therein). Conversely and according to our simulations,
these transition layers could be responsible for the stability of
fast, highly supersonic jets, preventing the mass-loading and
subsequent disruption. Finally, the type of shear layers developed by
models UST1/2 could mimic the transition layers invoked in models of
FRIs (Laing \& Bridle 2002a,b).

\section{Conclusions}
\label{sect:concl}

  We performed a number of simulations spanning a wide range of
parameters, such as Lorentz factor and specific internal energy, for a
general setup where a slab-sheared jet is perturbed with a set of
symmetric and antisymmetric sinusoidal perturbations, in order to
characterize the stability properties of relativistic jets.

  The most remarkable feature regarding the linear evolution of
instabilities is the finding of resonant modes in our simulations,
which were later confirmed by applying linear stability theory to
sheared flows. These modes are important for the long-term stability
properties of some jets (ST), for they remain collimated and unmixed,
thereby keeping a large amount of initial axial momentum. Jets in
which these modes do not grow fast enough with respect to longer
modes are disrupted either after a shock or by slow momentum
transfer and mixing.

  We classified jets depending on their nonlinear behaviour in three
different classes, which are clearly separated in a relativistic
internal Mach number vs. Lorentz factor plot
(Fig.~\ref{stabplane}). UST1 models are disrupted after a shock forms
in the early post-linear phase, and ambient gas penetrates deep into
the jet stream, decelerating and cooling the initial flow down. UST2
models are slowly decelerated by an efficient conversion of kinetic
energy into internal energy, which causes momentum transfer and
mixing. Finally, ST models present little expansion, but remain
collimated and isolated from the ambient by a hot shear layer. ST
models would fall into UST2, if resonant modes were not present, as
occurs for model D20 in Paper II.

Our simulations admit a clear interpretation in the context of the
morphological dichotomy of radio jets. Results presented here could
point to high Lorentz factor, highly supersonic jets as forming FRII
Class, whereas FRI jets would be related to slow, small Mach number
jets. In the former, the generation of a hot shear layer surrounding a
stable core could be related to the transversal structure observed in
several powerful jets.

\begin{acknowledgements}
  The authors want to thank J.A. Miralles for clarifying discussions
during the development of this work.  Calculations were performed on
the SGI Altix 3000 computer {\it CERCA} at the Servei d'Inform\`atica
de la Universitat de Val\`encia. This work was supported in part by
the Spanish Direcci\'on General de Ense\~nanza Superior under grant
AYA-2001-3490-C02 and by the Polish Committee for Scientific Research
(KBN) under grant PB~0656/P03/2004/26. M.H. acknowledges financial
support from the visitor program of the Universitat de
Val\`encia. M.P. benefited from a predoctoral fellowship of the
Universitat de Val\`encia ({\it V Segles} program).
\end{acknowledgements}

\end{document}